\documentclass[usenatbib]{mn2e}
\usepackage{graphicx,times}
\usepackage{bm,url}
\graphicspath{{./fig/}{./png/}}
\newcommand{\EQ}{\begin{equation}}
\newcommand{\EN}{\end{equation}}
\newcommand{\EQA}{\begin{eqnarray}}
\newcommand{\ENA}{\end{eqnarray}}
\newcommand{\nab}{\nabla}

\newcommand{\meanB}{\overline{B}}

\newcommand{\meanJ}{\overline{J}}

\newcommand{\meanemf}{\overline{\cal E} {}}
\newcommand{\meanEMF}{\overline{\mbox{\boldmath ${\mathcal E}$}} {}}

\newcommand{\meanBB}{\bm{\overline{{B}}}}
\newcommand{\meanJJ}{\bm{\overline{{J}}}}

\newcommand{\ppsi}{\bm{{{\psi}}}}
\newcommand{\uu}{\bm{{{u}}}}
\newcommand{\bb}{\bm{{{b}}}}
\newcommand{\BB}{\bm{{{B}}}}
\newcommand{\JJ}{\bm{{{J}}}}

\newcommand{\xx}{\bm{{{x}}}}

\newcommand{\RR}{\bm{{{R}}}}
\newcommand{\xxx}{\hat{\mbox{\boldmath $x$}} {}}
\newcommand{\yyy}{\hat{\mbox{\boldmath $y$}} {}}
\newcommand{\zzz}{\hat{\mbox{\boldmath $z$}} {}}
\newcommand{\x}{\times}
\newcommand{\p}{\partial}
\newcommand{\dd}{\mbox{d}}
\newcommand{\bzo}{{\bf 0}}
\newcommand{\ol}{\overline}

\def\ii{{\rm i}}

\def\Rm{R_\mathrm{m}}
\def\Rey{\mbox{\rm Re}}

\def\half{{\textstyle{1\over2}}}

\def\quarter{{\textstyle{1\over4}}}

\newcommand{\eq}[1]{(\ref{#1})}

\newcommand{\Eq}[1]{equation~(\ref{#1})}

\newcommand{\App}[1]{Appendix~\ref{#1}}
\newcommand{\Sec}[1]{Sect.~\ref{#1}}

\newcommand{\Fig}[1]{Fig.~\ref{#1}}

\newcommand{\Figs}[2]{Figs~\ref{#1} and \ref{#2}}

\title[Mean-field effects in the Galloway--Proctor flow]
{Mean-field effects in the Galloway--Proctor flow}
\author[K.-H. R\"adler and A.~Brandenburg]%
{Karl-Heinz R\"adler$^1$ and Axel Brandenburg$^2$
\thanks{E-mail: khraedler@arcor.de (KHR); brandenb@nordita.org (AB)}\\
$^{1}$ Astrophysical Institute Potsdam, An der Sternwarte 16, D-14482 Potsdam, Germany\\
$^2$NORDITA, Roslagstullsbacken 23, SE - 106 91 Stockholm, Sweden}
\date{}

\begin{document}

\pagerange{\pageref{firstpage}--\pageref{lastpage}} \pubyear{2008}

\maketitle

\begin{abstract}
The coefficients defining the mean electromotive force in a Galloway--Proctor flow are determined.
This flow shows a two-dimensional pattern and is helical.
The pattern wobbles in its plane.
Apart from one exception a circular motion of the flow pattern is assumed.
This corresponds to one of the cases considered recently by
Courvoisier, Hughes and Tobias (2006, Phys.\ Rev.\ Lett., 96, 034503).
An analytic theory of the $\alpha$ effect
and related effects in this flow is developed
within the second-order correlation approximation and a corresponding fourth-order approximation.
In the validity range of these approximations there is an $\alpha$ effect
but no $\gamma$ effect, or pumping effect.
Numerical results obtained with the test-field method,
which are independent of these approximations,
confirm the results for $\alpha$
and show that $\gamma$ is in general nonzero.
Both $\alpha$ and $\gamma$ show a complex dependency on the magnetic Reynolds number
and other parameters that define the flow,
that is, amplitude and frequency of the wobbling motion.
Some results for the magnetic diffusivity $\eta_{\rm t}$ and a related quantity
are given, too.
Finally a result for $\alpha$ in the case of a randomly varying
flow without the aforementioned
circular motion is presented.
This flow may be a more appropriate model for studying the $\alpha$ effect
and related effects in flows that are statistical isotropic in a plane.

\end{abstract}
\label{firstpage}
\begin{keywords}
magnetic fields --- MHD --- hydrodynamics -- turbulence
\end{keywords}

\section{Introduction}

In the astrophysical context, turbulent flows, e.g.\ in
stellar convection zones or in accretion discs and galaxies,
are generally anisotropic and time-dependent.
A simple model of a flow with such properties is that by \cite{GP92}.
This flow is two-dimensional, depends
only on two Cartesian coordinates, e.g.\ $x$ and $y$,
which can simplify the analysis significantly,
even in dynamo problems that are inherently three-dimensional.
The Galloway--Proctor (GP) flow is related to a flow considered by Roberts (1972).
The Roberts flow\footnote{As usual, the term Roberts flow refers to the
flow given by equation (5.1) of Roberts (1972).}
is an early example of a spatially periodic flow that produces an alpha effect.
The alpha term in the averaged form of the induction equation is crucial
to model the generation of large-scale magnetic fields from
small-scale helical fluid motions in stars and galaxies;
see, for example, \cite{Mof78}, \cite{Par79}, and \cite{KR80}
for standard references.
However, unlike the Roberts flow, the GP flow is time-dependent
with a flow pattern wobbling in
the $(x,y)$ plane in a circular fashion.
Both the GP flow and the Roberts flow have a velocity
component out of this plane
such that the flow can be fully helical, i.e.\
the velocity is proportional to its curl.

Particularly important is the dependence of the $\alpha$ effect on the
magnetic Reynolds number, $\Rm$.
While for the Roberts flow $\alpha$ declines with $\Rm$
in the large $\Rm$ limit,
in the case of the GP flow
according to the results by \cite{CHT06} (in the following referred to as CHT06) and \cite{Cou08}
there is a more complicated dependence on $\Rm$
with sign changes and no indication of convergence with increasing $\Rm$.

In many studies turbulent astrophysical flows have been modelled
by random forcing.
In the case of of helical isotropic turbulence such investigations show
that $\alpha$ approaches a finite value as soon as $\Rm$ exceeds a value of the order of unity.
This has been observed at least for Reynolds numbers up to 200 \citep{Sur_etal08}.

The purpose of this paper is to study the effects of the
GP flow in more detail in order to understand the influence
of time-dependence and anisotropy
on the value of $\alpha$ and other turbulent transport coefficients.
In particular, it is important to document the differences and similarities
with turbulent
flows that are statistically isotropic and irregular in space and time.
We focus attention here on the simplest case considered in CHT06 with a
flow being purely periodic in time
and add a few results for
a simple flow with random time dependence.

A number of similarities, but also some striking differences between
turbulent flows and the Roberts flow are known.
Similar in both flows is the fact that there is an $\alpha$ effect whose
magnitude increases with $\Rm$
as long as the latter does not exceed some value in the order of unity.
However, for larger $\Rm$, the $\alpha$ coefficient in isotropic turbulence
settles to a constant value
\citep{Sur_etal08}, while for the Roberts flow $\alpha$ tends to zero
as $\Rm \to \infty$
\citep{Soward87,Soward89,Radler_etal02a,Radler_etal02b}.
Furthermore, there is no $\gamma$ effect, or
pumping effect, neither for isotropic turbulence
nor for the Roberts flow.
On the other hand, in the time-dependent GP flow
$\gamma$ effects have been reported (CHT06).
A $\gamma$ effect corresponds to antisymmetric contributions of the
$\alpha$ tensor.
This raises the question about the possible existence of antisymmetric
contributions to the turbulent magnetic diffusivity tensor, or $\eta_{\rm t}$ tensor.

These aspects are now straightforward to address using the recently
developed test-field method to calculate numerically all components
of the $\alpha$ and $\eta_{\rm t}$ tensors defining the mean electromotive force $\meanEMF$
for a given flow field \citep{S05,S07}.
If, as we assume here, too, the mean magnetic field $\meanBB$ depends only on one
of the Cartesian coordinates, say $z$, only two $2 \x 2$ tensors for $\alpha$ and $\eta_{\rm t}$
are of interest.
The test-field method has recently been used to calculate diagonal and off-diagonal
components of $\eta_{\rm t}$ \citep{Betal08},
the magnetic Reynolds number dependence of $\alpha$ and $\eta_{\rm t}$
\citep{Sur_etal08}, as well as their scale dependence \citep{Betal08b}.
We begin by exploring general properties of the mean electromotive force in the GP flow
and present analytical results for coefficients like $\alpha$ and $\gamma$,
which are crucial for the electromotive force, gained in the second--order correlation approximation
and in a corresponding fourth--order approximation.
After explaining the test--field method we give a series of numerical results for
such coefficients, which are independent of approximations of that kind, and discuss them in detail.

\section{Mean--field electrodynamics with Galloway--Proctor flow}

\subsection{Definition of the problem}
\label{Definition}

Consider a magnetic field $\BB$ in an infinitely extended homogeneous conducting fluid
with constant magnetic diffusivity $\eta$ moving with a velocity $\uu$.
Its behavior is governed by
\EQ
\p_t \BB - \eta \nab^2 \BB - \nab \x (\uu \x \BB) = \bzo \, , \quad \nab \cdot \BB = 0 \, .
\label{eq01}
\EN
Referring to a Cartesian coordinate system $(x, y, z)$
the velocity $\uu$ is specified by
\EQ
\uu = - \zzz \times \nab \psi - \zzz \, k_{\rm H} \, \psi
\label{eq03}
\EN
with
\EQ
\psi = \frac{u_0}{k_{\rm H}} \, \big [\cos (k_{\rm H} x + \varphi_x)
    + \cos (k_{\rm H} y + \varphi_y) \big] \, .
\label{eq05}
\EN
Here $\zzz$ means the unit vector in the $z$ direction,
$k_{\rm H}$ is a positive constant such that $2 \pi /k_{\rm H}$ is the length of the diagonal of a flow cell,
and $\varphi_x$ and $\varphi_y$ are functions of time to be specified later.
Further we have $u_0 = u_{\rm rms}/\sqrt{2}$.
In the special case $\varphi_x = \varphi_y = 0$ the flow agrees with a Roberts flow.
For non-zero $\varphi_x$ or $\varphi_y$ a properly moving frame of reference can be found
in which we have again a steady Roberts flow pattern.
In our original frame each point of this pattern moves with the velocity
$- k_{\rm H}^{-1} ( \partial_t \varphi_x, \partial_t \varphi_y )$ in the $x y$ plane.
In \eq{eq03} the ratio of the flow components in the $xy$ plane and in $z$ direction
has been fixed such that the modulus of the average of the kinetic helicity
$\uu \cdot (\nab \x \uu)$ over all $x$ and $y$ for given $u_0$ takes its maximum.
With the signs chosen this average is equal to $- 2 u_0^2 k_{\rm H}$.

In view of the first example treated in CHT06 we specify the flow
generally defined by \eq{eq03} and \eq{eq05} further to be a Galloway-Proctor flow
and put
\EQ
\varphi_x = \epsilon \cos \omega t \, , \quad  \varphi_y = \epsilon \sin \omega t \, ,
\label{eq06}
\EN
where $\epsilon$ and $\omega$ are considered as non--negative constants.
We label this flow in what follows by (i).
Each point of this pattern moves with the frequency $\omega / 2 \pi$
on a circle with the radius $\epsilon / k_{\rm H}$.

To come closer to a turbulent situation CHT06 added a random function of time
to the arguments $\omega t$ in \eq{eq06}.
Another case of some interest occurs if we simply interpret $\varphi_x$ and $\varphi_y$
as random functions.
More precisely we put
\EQ
\varphi_x = \epsilon \phi_x (t/\tau_{\rm c}) \, , \quad
    \varphi_y = \epsilon \phi_y (t/\tau_{\rm c}) \, ,
\label{eq06b}
\EN
where $\epsilon$ is again a constant,
$\phi_x$ and $\phi_y$ are two independent but statistically equivalent random functions,
which take positive and negative values between $-1$ and $1$ and tend to zero
with growing moduli of the argument, and $\tau_{\rm c}$ is some correlation time.
We label this random flow by (ii).

\subsection{Mean--field concept}
\label{MeanFieldConcept}

Adopting the mean--field concept,
we denote mean fields by an overbar
and define them as averages over all $x$ and $y$.
We have then $\ol{\uu} = \bzo$.
Taking the average of \eq{eq01} we find
\EQ
\p_t \meanBB - \eta \nab^2 \meanBB - \nab \x \meanEMF = \bzo \, , \quad \nab \cdot \meanBB = 0 \, ,
\label{eq07}
\EN
with the mean electromotive force
\EQ
\meanEMF = \ol{\uu \x \bb} \, ,
\label{eq09}
\EN
where $\bb = \BB - \meanBB$.
[In \eq{eq07} $\nab$ reduces simply to $(0,0,\partial_z)$.]
From \eq{eq01} and \eq{eq07} we conclude that $\bb$ has to obey
\EQA
(\p_t - \eta \nab^2) \bb &=& (\meanBB \cdot \nab) \uu - (\uu \cdot \nab) \meanBB
\nonumber\\
&& +  \nab \x (\uu \x \bb - \ol{\uu \x \bb}) \, , \quad
    \nab \cdot \bb = 0 \, .
\label{eq11}
\ENA

We adopt here the assumption that the mean electromotive force $\meanEMF$ is,
apart from $\uu$ and $\eta$, completely determined by $\meanBB$ and its first spatial derivatives.
(This assumption will be relaxed in \Sec{Generalization}.)
This implies that there is no small--scale dynamo
and that sufficient time has elapsed since the initial instant
so that $\meanEMF$ no longer depends on any initial conditions.
Since $\meanBB$ is by definition independent of $x$ and $y$ its spatial derivatives
can be represented by $\nab \x \meanBB$.
We write simply $\meanJJ$ instead of $\nab \x \meanBB$,
being aware that the mean electric current density is really $\nab\times\meanBB/\mu$
(rather than $\meanJJ$),
where $\mu$ is magnetic permeability of the conducting fluid.
Clearly we have now $\meanJJ = (- \p \meanB_y /\p z, \p \meanB_x /\p z, 0)$.
For the sake of simplicity we
further assume that $\meanBB$ is steady.
In the so defined framework we may write
\EQ
\meanemf_i = \alpha_{ij} \meanB_j  - \eta_{ij} \meanJ_j
\label{eq13}
\EN
with tensors $\alpha_{ij}$ and $\eta_{ij}$ determined by $\uu$ and $\eta$ only.
Both $\alpha_{ij}$ and $\eta_{ij}$, and so $\meanemf_i$, too, depend in general on time.

We see from \eq{eq11} that, if $\meanBB$ is a uniform field,
$\bb$ and therefore $\meanEMF$ are independent of $\meanB_z$.
Hence we have $\alpha_{i3} = 0$.
Furthermore, since $\meanJ_z = 0$, clearly
$\eta_{i3}$ is without interest, and we put $\eta_{i3} = 0$.

\subsection{Mean electromotive force in case (i)}

For a more detailed investigation of $\meanEMF$ we focus on the fluid flow of type (i).
In this case $\alpha_{ij}$ and $\eta_{ij}$ are periodic in time with a basic period
equal to that of $\uu$, that is $2 \pi / \omega$,
or (as we will see below) a fraction of it.

Remarkably the velocity field $\uu = \uu (x,y,t)$ defined
by \eq{eq03}, \eq{eq05} and \eq{eq06}
is invariant under a $90^\circ$rotation about the $z$ axis and a simultaneous retarding
by $\pi / 2 \omega$ (that is, $\omega t \to \omega t - \pi/2$).
Consequently the $\alpha_{ij}$ in the correspondingly rotated coordinate system, which we denote
by $\alpha_{ij}'$, have to satisfy the relation
\EQ
\alpha_{ij}' (t - \pi / 2 \omega) = \alpha_{ij} (t) \, .
\label{eq15}
\EN
If we consider for a moment the change of the spatial coordinate system only
and ignore any time dependence we have
$\alpha_{11}' = \alpha_{22}$, $\alpha_{12}' = - \alpha_{21}$, $\alpha_{21}' = - \alpha_{12}$,
$\alpha_{22}' = \alpha_{11}$, $\alpha_{31}' = - \alpha_{32}$ and $\alpha_{32}' = \alpha_{31}$.
Hence \eq{eq15} provides us with
\EQA
\alpha_{11}(t) = + \alpha_{22}(t - \pi/2 \omega) , \; & \alpha_{22}(t) = + \alpha_{11}(t - \pi/2 \omega) ,
\nonumber\\
\alpha_{12}(t) = - \alpha_{21}(t - \pi/2 \omega) , \; & \alpha_{21}(t) = - \alpha_{12}(t - \pi/2 \omega) ,
\label{eq19}\\
\alpha_{31}(t) = - \alpha_{32}(t - \pi/2 \omega) , \; & \alpha_{32}(t) = + \alpha_{31}(t - \pi/2 \omega) .
\nonumber
\ENA
From the first two lines we conclude firstly that $\alpha_{11}$, $\alpha_{12}$, $\alpha_{21}$ and $\alpha_{22}$
have as functions of time a basic period of $\pi/\omega$ (not $2 \pi /\omega$)
and that $\alpha_{22} (t) = \alpha_{11} (t \pm \pi / 2 \omega)$
and $\alpha_{21} (t) = - \alpha_{12} (t \pm \pi / 2 \omega)$.
The last line of \eq{eq19} tells us that the averages of $\alpha_{31}$ and $\alpha_{32}$
over the period $2 \pi /\omega$ vanish so that $\alpha_{31}$ and $\alpha_{32}$
are simply oscillations around zero,
and that they change their signs under time shifts by $\pi /\omega$.
Our reasoning for $\alpha_{ij}$ applies analogously to $\eta_{ij}$.

We write down the result of these considerations in the form
\EQA
& \alpha_{11} = \tilde{\alpha} (t) \quad   & \alpha_{22} = \tilde{\alpha} (t - \pi/2 \omega)
\nonumber\\
& \alpha_{12} = - \tilde{\gamma} (t) \quad  & \alpha_{21} = \tilde{\gamma} (t - \pi/2 \omega)
\nonumber\\
& \alpha_{31} = \tilde{\kappa} (t) \quad  & \alpha_{32} = \tilde{\kappa} (t - \pi/2 \omega)
\nonumber\\
& \eta_{11} = \tilde{\eta}_{\rm t} (t) \quad  & \eta_{22} = \tilde{\eta}_{\rm t} (t - \pi/2 \omega)
\label{eq21}\\
& \eta_{12} =  - \tilde{\delta} (t) \quad  & \eta_{21} = \tilde{\delta} (t - \pi/2 \omega)
\nonumber\\
& \eta_{31} =  \tilde{\lambda} (t) \quad  & \eta_{32} = \tilde{\lambda} (t - \pi/2 \omega)
\nonumber
\ENA
Here $\tilde{\alpha}$, $\tilde{\gamma}$, $\tilde{\eta}_{\rm t}$ and $\tilde{\delta}$
are in general periodic functions of time with the basic period $\pi/\omega$,
but $\tilde{\kappa}$ and $\tilde{\lambda}$ are periodic functions with period $2 \pi / \omega$,
which show sign changes under any time shift by $\pi / \omega$
and vanish under averaging over the period $2 \pi / \omega$.

From \eq{eq13} and \eq{eq21} we conclude
\EQA
\meanEMF &=& \tilde{\alpha} (\xxx \cdot \meanBB) \xxx + \tilde{\alpha}^\dag (\yyy \cdot \meanBB) \yyy
\nonumber\\
&&    - \tilde{\gamma} (\yyy \cdot \meanBB) \xxx + \tilde{\gamma}^\dag (\xxx \cdot \meanBB) \yyy
\nonumber\\
&&   - \tilde{\eta}_{\rm t} (\xxx \cdot \meanJJ) \xxx - \tilde{\eta}^\dag_{\rm t} (\yyy \cdot \meanJJ) \yyy
\label{eq22}\\
&&   + \tilde{\delta} (\yyy \cdot \meanJJ) \xxx - \tilde{\delta}^\dag (\xxx \cdot \meanJJ) \yyy
\nonumber\\
&& + \left[\tilde{\kappa} (\xxx \cdot \meanBB) + \tilde{\kappa}^\dag (\yyy \cdot \meanBB)
     + \tilde{\lambda} (\xxx \cdot \meanJJ) + \tilde{\lambda}^\dag (\yyy \cdot \meanJJ) \right] \zzz \, .
\nonumber
\ENA
Here $\tilde{\alpha}^\dag$, $\tilde{\gamma}^\dag$,  $\tilde{\eta}^\dag_{\rm t}$ and $\tilde{\delta}^\dag$
differ only by a phase shift of $\pi / 2$ from $\tilde{\alpha}$, $\tilde{\gamma}$, $\tilde{\eta}_{\rm t}$
and $\tilde{\delta}$, respectively,
and $\tilde{\kappa}^\dag$ and $\tilde{\lambda}^\dag$ by a phase shift of $\pi$
from $\tilde{\kappa}$ and $\tilde{\lambda}$.

In addition to fields as defined above by averaging over $x$ and $y$ we consider also
time-averaged mean fields defined by additional averaging
over a time interval of length $2 \pi/\omega$
(but we refer to them only if explicitly indicated).
When speaking of time averaging in what follows we always refer to this interval.
For time-averaged mean fields \eq{eq22} turns into
\EQ
\meanEMF = \alpha \left[ \meanBB - (\zzz \cdot \meanBB) \zzz \right]
    + \gamma \zzz \x \meanBB - \eta_{\rm t} \meanJJ - \delta \zzz \x \meanJJ \, ,
\label{eq23}
\EN
where $\alpha$, $\gamma$, $\eta_{\rm t}$ and $\delta$
are time averages of $\tilde{\alpha}$, $\tilde{\gamma}$, $\tilde{\eta}_{\rm t}$ and $\tilde{\delta}$.%
\footnote{In view of the signs of $\alpha$ and $\gamma$ we deviate here from representations as given, e.g., in
R\"adler et al.\ (2002a) but follow CHT06.}

In the special case of the Roberts flow,
i.e.\ $\epsilon = 0$, the coefficient
$\tilde{\alpha}$ is independent of time and so coincides with
$\tilde{\alpha}^\dag$,
and this applies analogously to $\tilde{\gamma}$, $\tilde{\eta}$, $\tilde{\delta}$,
$\tilde{\kappa}$ and $\tilde{\lambda}$.
In addition in this case the inversion of $\zzz$ in \eq{eq03}
is equivalent to a shift of the flow pattern,
e.g., by $\pi / \sqrt{2} \, k_{\rm H}$ along $y = x$.
Since such a
shift does not change averages, $\meanEMF$ as given by \eq{eq23} must be even in $\zzz$.
Therefore we have then $\gamma = \delta = 0$.
Nonzero $\gamma$ and $\delta$ terms in \eq{eq23} require a break
of this symmetry, that is, a preference of $\zzz$ over $-\zzz$,
and this may occur as a consequence of the aforementioned circular motion of the flow pattern.

We override for a moment our restriction to non-negative values of the frequency $\omega$
and
admit also negative ones.
For $\omega > 0$ the circular motion of the flow pattern defines,
together with the $z$ direction, a right--handed screw, and for $\omega < 0$ a left--handed one.
We conclude from this fact that inversion of the sign of $\omega$ has no other consequences
than inversion of the signs of $\gamma$ and $\delta$.

\subsubsection*{Second--order approximation}
\label{SOCA}

The task of determination of $\meanEMF$ is now reduced to the determination
of the six functions $\tilde{\alpha}$, $\tilde{\gamma}$,
$\tilde{\eta}_{\rm t}$, $\tilde{\delta}$, $\tilde{\kappa}$ and $\tilde{\lambda}$
which occur in \eq{eq21} and \eq{eq22}.
As a first step in that direction we investigate $\meanEMF$
within the second--order correlation approximation (SOCA).
Later we will proceed to a corresponding fourth--order approximation.

SOCA is defined by the neglect of the term with $\uu \x \bb - \ol{\uu \x \bb}$
on the right--hand side of equation \eq{eq11} for $\bb$, which turns so into
\EQ
(\p_t - \eta \nab^2) \bb = (\meanBB \cdot \nab) \uu  - (\uu \cdot \nab) \meanBB \, , \quad
    \nab \cdot \bb = 0 \, .
\label{eq31}
\EN
We may solve this equation with $\uu$ as given by \eq{eq03} and \eq{eq05} analytically
and calculate then $\meanEMF$, see \App{ApSOCA}.
When choosing the form \eq{eq22} of the result we have
\EQA
\tilde{\alpha} \!\!\!&=&\!\!\! u_0 \Rm  \chi^{(2)} (t) \, ,
\nonumber\\
\tilde{\eta}_{\rm t} \!\!\!&=&\!\!\! \half u_0 k_{\rm H}^{-1} \Rm
    \left[ \chi^{(2)} (t) + \chi^{(2)} (t - \pi / 2 \omega) \right] \, ,
\label{eq33}\\
\tilde{\gamma} \!\!\!&=&\!\!\! \tilde{\delta} = \tilde{\kappa} = \tilde{\lambda} = 0 \, .
\nonumber
\ENA
Here we have used the definition
\EQ
R_m = u_0 / \eta k_{\rm H} \, ,
\label{eq35}
\EN
and $\chi^{(2)}$ is given by
\EQ
\chi^{(2)} (t) = \int_0^\infty \mbox{CC} (\omega t, \omega t - q \tau )
\, \mbox{e}^{-\tau} \, \dd \tau
\label{eq36}
\EN
where
\EQ
\mbox{CC} (a, b) = \cos \left[ \epsilon (\cos a - \cos b) \right]
\label{eq37}
\EN
and
\EQ
q = \omega / \eta k_{\rm H}^2 \, .
\label{eq39}
\EN
The parameter $q$ gives, apart from a factor $2 \pi$,
the ratio of the decay time of a magnetic structure
with a length scale $2 \pi / k_{\rm H}$,
that is $(2 \pi)^2 / \eta k_{\rm H}^2$,
and the wobble period $2 \pi / \omega$ of the flow pattern.
In the case of small $q$ the magnetic field follows the fluid motion immediately,
but for large $q$ it does so only with large delay.
These two cases are sometimes labelled as ``low conductivity limit" and ``high conductivity limit",
respectively.

In agreement with the general findings summarized in \eq{eq21},
the function $\chi^{(2)}$ is periodic in time with a basic period $\pi / \omega$.
Whereas $\tilde{\alpha}$ and $\tilde{\alpha}^\dag$ differ by a phase shift of $\pi/2$,
$\tilde{\eta}_{\rm t}$ and $\tilde{\eta}_{\rm t}^\dag$ coincide.
$\chi^{(2)}$ satisfies $|\chi^{(2)}| \leq 1$.
It must be positive as long as $\epsilon \leq \pi/4$
but may otherwise take negative values, too.
If $\epsilon = 0$, or $\epsilon \not= 0$ and $q = 0$
(what corresponds to the low--conductivity limit),
$\chi^{(2)}$ is independent of time and equal to unity.
In \App{ApSOCA} some numerically determined values of $\chi^{(2)}$ are given.
We note further that
\EQ
\chi^{(2)} (t) = 1 - (\epsilon q)^2 \sin^2 \omega t \quad \mbox{if} \quad
    \epsilon q \ll 1 \;\; \mbox{and} \;\; q \ll 1 \, ,
\label{eq41}
\EN
and
\EQ
\chi^{(2)} (t) \to 0 \quad \mbox{as} \;\; q \to \infty \, .
\label{eq42}
\EN

For time--averaged mean fields we have again \eq{eq23},
now with
\EQ
\alpha = u_0 R_m \chi^{(2)}_0 \, , \quad \eta_{\rm t} = u_0 k_{\rm H}^{-1} \Rm \chi^{(2)}_0 \, , \quad
    \gamma = \delta = 0 \, ,
\label{eq43}
\EN
where $\chi^{(2)}_0$ means the time average of $\chi^{(2)}$ over the period $\pi/\omega$.
We point out that the time average of a function of $\omega t$,
say $f (\omega t)$, over an interval of the length $\pi/\omega$ is independent of $\omega$.
This is obvious from
$(\omega/\pi) \int_0^{\pi/\omega} f (\omega t) \, \dd t = (1/\pi) \int_0^\pi f (\varphi) \, \dd \varphi$.
Hence $\chi^{(2)}_0$ does not explicitly, but only in a indirect way via $q$, depend on $\omega$.

If $\epsilon = 0$, or $\epsilon \not= 0$ and $q = 0$, we have $\chi^{(2)}_0 = 1$.
For $\epsilon = 0$ we fall back to the Roberts flow.
Indeed, the result \eq{eq43} with $\chi^{(2)}_0 = 1$ agrees with
earlier results for this flow; see \App{ApRob}.
In view of \eq{eq43} we note further
\EQ
\chi^{(2)}_0 =  1 -  \half (\epsilon q)^2 \quad \mbox{if} \quad \epsilon q \ll 1 \;\; \mbox{and} \;\; q \ll 1
\label{eq45}
\EN
and
\EQ
\chi^{(2)}_0 \to 0 \quad \mbox{as} \quad q \to \infty \, .
\label{eq46}
\EN
The last statement implies $\alpha / u_0 \Rm \to 0$ as $q \to \infty$.

As we know from general considerations on SOCA (e.g., Krause \& R\"adler 1980)
the range of applicability of SOCA depends on $q$.
For small $q$ a sufficient condition for its validity reads $\Rm \ll 1$.
For large $q$ such a condition is $\Rm / q \ll 1$.

\subsubsection*{Higher--order approximations}
\label{HOCA}

Going now beyond SOCA we start again with Eq.~(\ref{eq11}) for $\bb$ and put
\EQ
\bb = \bb^{(1)} + \bb^{(2)} + \bb^{(3)} + \cdots
\label{eq53}
\EN
with $\bb^{(n)}$ being of the order $n$ in $\uu$, and correspondingly
\EQ
\meanEMF = \meanEMF^{(2)} + \meanEMF^{(3)} + \meanEMF^{(4)} + \cdots \, ,  \;\;
    \meanEMF^{(n+1)} = \langle \uu \x \bb^{(n)} \rangle \, , \; n \geq 1 \, .
\label{eq55}
\EN
In that sense $\bb$ and $\meanEMF$ in \Sec{SOCA}
have to be interpreted as $\bb^{(1)}$
and $\meanEMF^{(2)}$, respectively.

From \eq{eq11} and \eq{eq53} we obtain \eq{eq31}, now with $\bb^{(1)}$ instead of $\bb$, and further
\EQA
(\p_t - \eta \nab^2) \bb^{(n+1)} \!\! &=& \!\! (\bb^{(n)} \! \cdot \nab) \uu - (\uu \cdot \nab) \bb^{(n)} \!
\nonumber\\
&& \qquad \qquad \qquad   - \nab  \x (\ol{\uu \x \bb^{(n)}}) \, ,
\label{eq59}\\
&& \qquad   \nab \cdot \bb^{(n+1)} = 0 \, , \quad n \geq 1 \, .
\nonumber
\ENA

Using our
result for $\bb^{(1)}$ and (\ref{eq59}) we have calculated $\bb^{(2)}$.
It turns out that the average of $\uu \x \bb^{(2)}$ vanishes, that is $\meanEMF^{(3)} = \bzo$.
In the same way we may calculate $\bb^{(3)}$ and $\meanEMF^{(4)}$.
However, these calculations are rather tedious.
For the sake of simplicity we have ignored all contributions to $\meanEMF^{(4)}$
resulting from derivatives of $\meanBB$, that is, the terms with $\tilde{\eta}_{\rm t}$,
$\tilde{\delta}$ and $\tilde{\lambda}$ in \eq{eq22}.
Some details of the calculations are explained in \App{ApHOCA}.

Considering the results of all approximations up to the fourth order and referring again to \eq{eq22}
we have now
\EQA
\tilde{\alpha} \!\! &=& \!\! u_0 \Rm \left[ \chi^{(2)} (t) - \half \Rm^2 \chi^{(4 \, \alpha)} (t) \right] \, ,
\nonumber\\
\tilde{\gamma} \!\! &=& \!\! \half u_0 \Rm^3 \chi^{(4 \, \gamma)} (t) \, , \quad \tilde{\kappa} = 0 \, .
\label{eq63}
\ENA
The functions $\chi^{(4 \, \alpha)}$ and $\chi^{(4 \, \gamma)}$ are given by
\EQA
\chi^{(4 \alpha)} \!\!\!&=&\!\!\! 2 \int_0^\infty \!\! \int_0^\infty \!\! \int_0^\infty
    \mbox{CC} (\omega t, \omega t - q (\tau' + \tau'' + \tau'''))
\nonumber\\
&&  \qquad \qquad \quad \mbox{CS} (\omega t - q \tau', \omega t - q (\tau' + \tau''))
\nonumber\\
&&  \qquad \qquad \quad \exp (- (\tau'+ 2 \tau'' + \tau''') \, \dd \tau' \, \dd \tau'' \, \dd \tau''' \, ,
\nonumber\\
\chi^{(4 \gamma)} \!\!\!&=&\!\!\! 2 \int_0^\infty \!\! \int_0^\infty \!\! \int_0^\infty
     \mbox{SC} (\omega t, \omega t - q (\tau' + \tau''))
\label{eq65}\\
&&  \qquad \qquad \quad \mbox{SS} (\omega t - q \tau', \omega t - q (\tau' + \tau'' + \tau'''))
\nonumber\\
&&  \qquad \qquad \quad \exp (- (\tau'+ 2 \tau'' + \tau''') \, \dd \tau' \, \dd \tau'' \, \dd \tau''' \, ,
\nonumber
\ENA
with $\mbox{CC}$ as defined by \eq{eq37} and analogously defined quantities $\mbox{CS}$, $\mbox{SC}$ and $\mbox{SS}$,
\EQA
\mbox{CS} \, (a, b) \!\!\!&=&\!\!\! \cos \left[\epsilon (\sin a - \sin b) \right] \, ,
\nonumber\\
\mbox{SC} \, (a, b) \!\!\!&=&\!\!\! \sin \left[\epsilon (\cos a - \cos b) \right] \, ,
\label{eq67}\\
\mbox{SS} \, (a, b) \!\!\!&=&\!\!\! \sin \left[\epsilon (\sin a - \sin b) \right] \, .
\nonumber
\ENA
Note that $\mbox{CC}$ and $\mbox{CS}$ are symmetric but $\mbox{SC}$ and $\mbox{SS}$ antisymmetric in the two arguments.

Like $\chi^{(2)}$ both $\chi^{(4 \alpha)}$ and $\chi^{(4 \gamma)}$ oscillate with a basic period $\pi/\omega$.
They satisfy $|\chi^{(4 \alpha)}| \leq 1$ and $|\chi^{(4 \gamma)}| \leq 1$.
Further $\chi^{(4 \alpha)}$ is positive as long as $|\epsilon| < \pi/4$.
In contrast to $\chi^{(4 \alpha)}$, however, the time average of $\chi^{(4 \gamma)}$
over a period $\pi/\omega$ is equal to zero.
Whereas $\chi^{(4 \alpha)}$ is even, $\chi^{(4 \gamma)}$ is odd in $\omega$.
We have further
\EQA
\chi^{(4 \alpha)} \!\!\!&=&\!\!\! 1 - \quarter (\epsilon q)^2 (1 + 16 \sin^2 \omega t ) \, ,
\nonumber\\
\chi^{(4 \gamma)} \!\!\!&=&\!\!\! \textstyle{5 \over 2} (\epsilon q)^2 \sin \omega t \, \cos \omega t \, , \quad
     \mbox{if} \;\; \epsilon q \ll 1 \;\; \mbox{and} \;\; q \ll 1 \, ,
\label{eq69}
\ENA
and
\EQ
\chi^{(4 \alpha)}\, ,  \chi^{(4 \gamma)} \to 0 \quad \mbox{as} \quad q \to \infty \, .
\label{eq70}
\EN

For time--averaged mean fields again relation \eq{eq23} applies, now with
\EQ
\alpha = u_0 \Rm \left[ \chi^{(2)}_0 - \half \Rm^2 \chi^{(4 \alpha)}_0 \right] \, , \quad \gamma = 0 \, ,
\label{eq71}
\EN
where $\chi^{(4 \alpha)}_0$ is the time average of $\chi^{(4 \alpha)}$.
Like $\chi^{(2)}_0$ also $\chi^{(4 \alpha)}_0$ does not explicitly depend on $\omega$.
Unfortunately, values for $\eta_{\rm t}$ and $\delta$ are not available.
We have
\EQA
\chi^{(4 \alpha)}_0 = 1 - \textstyle{9 \over 4} (\epsilon q)^2  \, , \quad
    \chi^{(4 \gamma)}_0 = 0 \, , \quad \mbox{if} \;\; \epsilon q \ll 1 \; \; \mbox{and} \;\; q \ll 1 \, ,
\label{eq73}
\ENA
and
\EQ
\chi^{(4 \alpha)}_0, \chi^{(4 \gamma)}_0  \to 0 \quad \mbox{as} \quad q \to \infty \, .
\label{eq73b}
\EN
With \eq{eq71} we find then
\EQA
\alpha \!\!\!&=&\!\!\! u_0 \Rm \left[ 1 - \half (\epsilon q)^2
    - \half \Rm^2 ( 1 - \textstyle {9 \over 4}(\epsilon q)^2) \right]
\nonumber\\
&& \qquad \qquad \qquad \qquad \qquad \quad \mbox{if} \;\; \epsilon q \ll 1 \; \; \mbox{and} \;\; q \ll 1\, .
\label{eq75}
\ENA

Results of higher approximations are very desirable but require heavy efforts.
We suspect that in the approximation of sixth order in $\uu$ the time averages
of $\tilde{\gamma}$ and $\tilde{\delta}$,
and so the coefficients $\gamma$ and $\delta$ in \eq{eq23} no longer vanishes.
This presumption is supported by numerical results (see below).

\subsection{Mean electromotive force in case (ii)}
\label{Randomflow}

Modifying the considerations on case (i) properly we may conclude that relation \eq{eq23},
again considered for time-averaged fields, applies for the fluid flow of type (ii)
with $\gamma = \delta = 0$.
By contrast to case (i) the correlation between velocity components at different times
vanishes if the time difference becomes very large.

\begin{figure}\begin{center}
\includegraphics[width=\columnwidth]{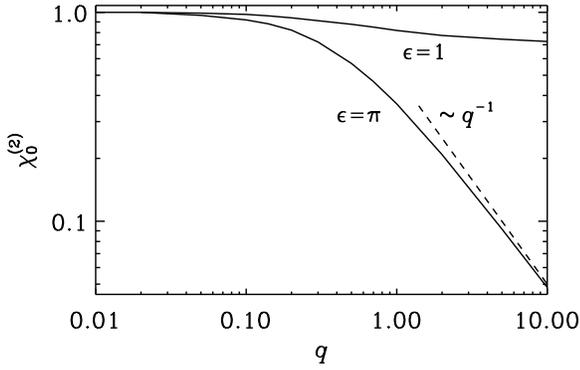}
\end{center}\caption[]{
Dependence of $\chi^{(2)}_0$ on $q$ for two different $\epsilon$,
calculated numerically on the basis of \Eq{eq137}.
The dashed line shows that $\chi^{(2)}_0$ with $\epsilon = \pi$ behaves like $q^{-1}$ for large $q$.
}\label{pintegral_cc_random}\end{figure}

Modifying also the SOCA calculations described above and in \App{ApSOCA}
correspondingly we find again \eq{eq43},
but with $\chi^{(2)}_0$ being the time average of
\EQ
\chi^{(2)}(t) = \int_0^\infty \! \cos \big\{ \epsilon [ \phi_x (t/\tau_{\rm c}) - \phi_x (t/\tau_{\rm c} - q \tau )] \big\}
     \, \mbox{e}^{-\tau} \, \dd \tau \, ,
\label{eq137}
\EN
where $q$ is now defined by
\EQ
q = (\tau_{\rm c} \eta k_{\rm H}^2)^{-1} \, .
\label{eq139}
\EN
We have here again $\chi^{(2)} = 1$ for $q = 0$,
and $\chi^{(2)}$ vanishes for $q \to \infty$.
Like $\chi^{(2)}$ also $\chi^{(2)}_0$ depends on $\epsilon$ and $q$ but no longer
explicitly on $\tau_{\rm c}$.
We have calculated $\chi^{(2)}_0$ on the basis of \Eq{eq137} under the assumption
that $\phi_x$ is always constant over time intervals of a given length.
\Fig{pintegral_cc_random} shows dependencies on $\epsilon$ and $q$.

\section{Test-field method}
\label{tfmeth}

We will determine
numerically the elements of the tensors $\alpha_{ij}$ and $\eta_{ij}$
introduced with \eq{eq13}, but with $1 \leq i, j \leq 2$ only,
employing the test-field method of \cite{S05,S07}.

We will calculate $\meanEMF = \ol{\uu \x \bb}$ from numerical solutions $\bb$ of \eq{eq11},
with $\meanBB$ replaced by one out of four test fields $\meanBB^{pq}$,
\EQA
\meanBB^{1 {\rm c}} &=& B \, (\cos kz, 0, 0) \, , \quad \meanBB^{2 {\rm c}} = B \, ( 0, \cos kz, 0) \, ,
\nonumber\\
\meanBB^{1 {\rm s}} &=& B \, (\sin kz, 0, 0) \, , \quad \meanBB^{2 {\rm s}} = B \, ( 0, \sin kz, 0) \, ,
\label{eq103}
\ENA
where $B$ and $k$ are a constants.
Repeating this for all test fields, denoting the $\meanEMF$ that belongs
to a given $\meanBB^{pq}$ by $\meanEMF^{pq}$, and using \eq{eq13} we find
\EQA
\meanemf_i^{p {\rm c}} (z) \!\!\! &=& \!\!\! B \left(\alpha_{ip} \cos kz - \eta^\dag_{ip} k \sin kz \right) \, ,
\nonumber\\
\meanemf_i^{p {\rm s}} (z) \!\!\! &=& \!\!\! B \left(\alpha_{ip} \sin kz + \eta^\dag_{ip} k \cos kz \right) \, ,
\label{eq103b}
\ENA
for $1 \leq i, j \leq 2$, where
\EQ
\eta^\dag_{ip} = \eta_{il} \epsilon_{lp3}
    = \pmatrix{- \eta_{12} & \eta_{11} \cr - \eta_{22} & \eta_{21}} \, .
\label{eq105}
\EN
From this we conclude
\EQA
\alpha_{ij} &=&  B^{-1} \left[ \meanemf_i^{j \, \rm{c}} (z) \cos k z
    + \meanemf_i^{j \, \rm{s}} (z) \sin k z \right] \, ,
\nonumber\\
\eta^\dag_{ij} &=& - (k B)^{-1} \left[ \meanemf_i^{j \, \rm{c}} (z) \sin k z
    - \meanemf_i^{j \, \rm{s}} (z) \cos k z \right] \, ,
\label{eq107}
\ENA
again for $1 \leq i, j \leq 2$.

We point out that, although the $\meanEMF^{pq}$ depend on $z$,
the $\alpha_{ij}$ and $\eta_{ij}$ have to be independent of $z$.
We further note that relation \eq{eq13}, on which these considerations are based,
can only be justified under the assumption that all higher than first--order
spatial derivatives of $\meanBB$ are negligible.
The derivatives of order $n$ of our test fields $\meanBB^{pq}$ are proportional to $k^n$.
For this reason the results \eq{eq107} apply in a strict sense only in the limit $k \to 0$
\citep[cf.][]{Betal08b}.

Let us focus here on case (i).
After having calculated the $\alpha_{ij}$ and $\eta_{ij}$ in the way indicated above
we may determine the $\tilde{\alpha}$, $\tilde{\gamma}$, $\tilde{\eta}_{\rm t}$ and $\tilde{\delta}$
according to \eq{eq21}, that is,
\EQA
\tilde{\alpha} (t) &=& \half \left[ \alpha_{11} (t)+ \alpha_{22} (t + \pi /2 \omega) \right] \, ,
\nonumber\\
\tilde{\gamma} (t) &=& - \half \left[ \alpha_{12} (t) - \alpha_{21} (t + \pi /2 \omega) \right] \, ,
\nonumber\\
\tilde{\eta}_{\rm t} (t) &=& \half \left[ \eta_{11} (t) + \eta_{22} (t + \pi /2 \omega) \right] \, ,
\label{eq109}\\
\tilde{\delta} (t) &=& - \half \left[ \eta_{12} (t) - \eta_{21} (t + \pi /2 \omega) \right] \, .
\nonumber
\ENA
We are, however, mainly interested in the time-independent coefficients
$\alpha$, $\gamma$, $\eta$ and $\delta$ that are relevant for time--averaged mean fields
as addressed in \eq{eq23}.
They are just time averages of the $\tilde{\alpha}$, $\tilde{\gamma}$, $\tilde{\eta}_{\rm t}$
and $\tilde{\delta}$, that is
\EQA
\alpha \!\!\! &= \half \big( \langle \alpha_{11} \rangle + \langle \alpha_{22} \rangle \big) \, , \quad
\gamma \!\!\! &= - \half \big( \langle \alpha_{12} \rangle - \langle \alpha_{21} \rangle \big)
\nonumber\\
\eta_{\rm t} \!\!\! &= \half \big( \langle \eta_{11} \rangle + \langle \eta_{22} \rangle \big) \, , \quad
\delta \!\!\! &= - \half \big( \langle \eta_{12} \rangle - \langle \eta_{21} \rangle \big) \, ,
\label{eq109b}
\ENA
where $\langle \cdots \rangle$ means averaging over a time interval of length $\pi / \omega$.
In case (ii) the relations \eq{eq109b} apply with $\langle \cdots \rangle$ interpreted
as averaging over a sufficiently long time.

\section{A Generalization}
\label{Generalization}

So far we have assumed that the mean electromotive force $\meanEMF$ in a given point
is completely determined by $\meanBB$ and its first spatial derivatives in this point.
If we relax this assumption we may proceed as in \cite{Betal08b}.
In that sense we may replace \eq{eq13},
applied to time--averaged mean fields, by
\EQ
\meanemf_i (z) = \int \big[ \hat{\alpha}_{ij} (\zeta) \meanB_j (z - \zeta)
    - \hat{\eta}_{ij} (\zeta) \meanJ_j (z - \zeta) \big] \, \dd \zeta
\label{eqxx01}
\EN
with kernels $\hat{\alpha}_{ij}$ and $\hat{\eta}_{ij}$.
When
using a Fourier transformation $Q (z) = \int \tilde{Q} \exp (\mbox{i} k z) \dd z$,
this turns into
\EQ
\tilde{\meanemf}_i (k) = \tilde{\alpha}_{ij} (k) \tilde{\meanB}_j (k)
    - \tilde{\eta}_{ij} (k) \tilde{\meanJ}_j (k) \, ,
\label{eqxx03}
\EN
where
\EQ
\tilde{\alpha}_{ij} (k) = \!\! \int \! \hat{\alpha}_{ij} (\zeta) \cos k \zeta \, \dd \zeta \, , \;\;
     \tilde{\eta}_{ij} (k) = \!\! \int \! \hat{\eta}_{ij} (\zeta) \cos k \zeta \, \dd \zeta \, .
\label{eqxx05}
\EN
In this understanding
the relations \eq{eq103b}--\eq{eq107} apply with $\alpha_{ij}$ and $\eta_{ij}$ being
replaced by $\tilde{\alpha}_{ij}$ and $\tilde{\eta}_{ij}$, which have a well--defined meaning for all $k$
(not only in the limit $k \to \infty$).

\section{Results}
\label{Results}

\subsection{Units and dimensionless parameters}

It is appropriate to give
$\tilde{\alpha}$ and $\tilde{\gamma}$ as well as $\alpha$ and $\gamma$
in units of $u_0$, and
$\tilde{\eta}_{\rm t}$, $\tilde{\delta}$, $\eta_{\rm t}$ and $\delta$
in units of $u_0 / k_{\rm H}$.
The remaining dimensionless parts of these coefficients are then,
apart from the time dependencies of $\tilde{\alpha}$, $\tilde{\gamma}$, $\tilde{\eta}_{\rm t}$
and $\tilde{\delta}$,
functions of the dimensionless parameters $\Rm$, $\epsilon$ and $q$
introduced through
\eq{eq35}, \eq{eq05}, and either \eq{eq39} or \eq{eq139}.
Instead of $q$ we may also use the dimensionless quantity $\tilde{\omega}$
defined by
\EQ
\tilde{\omega} = q / \Rm \, .
\label{eq111}
\EN
In case (i) we have so $\tilde{\omega} = \omega / u_0 k_{\rm H}$,
which is the ratio of the turnover time $(u_0 k_{\rm H} / 2 \pi)^{-1}$ to the wobble period $2 \pi / \omega$.
In case (ii) applies $\tilde{\omega} = (\tau_{\rm c} u_0 k_{\rm H})^{-1}$, and this is,
apart from a factor $2 \pi$, the ratio of that turnover time to the time $\tau_{\rm c}$
introduced with the random flow.

\subsection{Case (i)}
\label{comp}

\subsubsection*{Comparison with CHT06}

We show first that our method reproduces results by CHT06.
We suppose that our $\Rm$ is related to the magnetic Reynolds number, say $\Rm^{\rm CHT}$,
used but not explicitly defined there, by $\Rm = \sqrt{3/2} \, \Rm^{\rm CHT}$.
While in CHT06 dependencies of the results on $\Rm$ and $\epsilon$ are considered,
no values of $q$ or $\tilde{\omega}$ are given.
We suppose that the calculations have actually been carried out with $\tilde{\omega} = \sqrt{2/3}$.
Finally we suppose that the unit of $\alpha$ and $\gamma$ used by CHT06 is $\omega/k_{\rm H}$.

\begin{figure}\begin{center}
\includegraphics[width=\columnwidth]{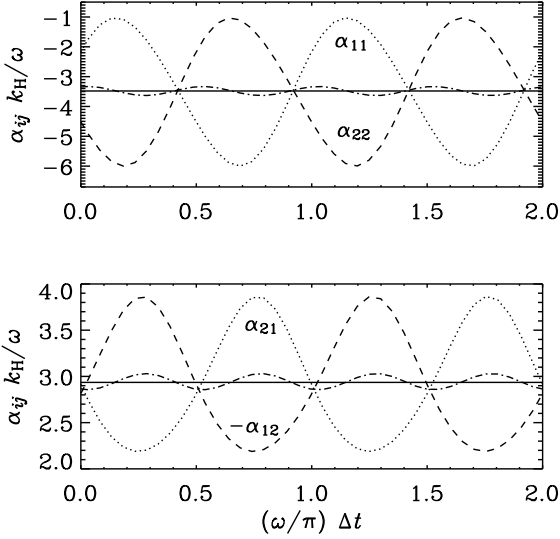}
\end{center}\caption[]{
Time dependence of $\alpha_{ij}$ for the parameters used
in Fig.~1 of CHT06 which, in our normalization, are
$\Rm=78$, $\epsilon=3/4$, and
$\tilde{\omega} = \sqrt{2/3}$.
Here, $\Delta t=t-t_0$, where
$t_0 =300 / \omega$
is the final time shown in Fig.~1 of CHT06.
The dotted lines refer to $\alpha_{11}$ and $\alpha_{21}$, respectively,
the dashed lines to $\alpha_{12}$ and $\alpha_{22}$,
and the dash-dotted lines to
$(\alpha_{11}+\alpha_{22})/2$ and $(\alpha_{21}-\alpha_{12})/2$.
The straight solid lines give
the time averages of the latter quantities,
that is, $\alpha$ and $\gamma$.
}\label{palpt}\end{figure}

With a view to Fig.~1 of CHT06 we have
carried out calculations with
$\Rm = \sqrt{3/2} \x 64 \approx 78$, $\epsilon = 3/4$
and $\tilde{\omega} = \sqrt{2/3}$.
Our results for the $\alpha_{ij}$ obtained with these parameters and given in this particular case
in units of $\omega / k_{\rm H}$ are presented in our \Fig{palpt}.
We see in particular that $\alpha_{11}$ and $\alpha_{22}$ vary between $-6$ and $-1$
with a period $\pi / \omega$.
As far as $\alpha_{11}$ is concerned this agrees with the result for $\langle \uu \times \bb \rangle_x$
shown in Fig.~1 of CHT06.
Also the initial evolution of $\alpha_{11}$, which is not shown here, agrees with this figure.
Furthermore, in our \Fig{palpt}
the phase shift by $\pi /2$ between $\alpha_{11}$ and $\alpha_{22}$
discussed in \Sec{MeanFieldConcept} is clearly visible.
Our results for $\alpha_{12}$ and $\alpha_{21}$ lead to a value of $\gamma$,
which agrees in modulus but differs in sign from that of CHT06.
(To obtain their sign we need to replace $\omega$ by $-\omega$.)
With the above values of $\Rm$ and $\tilde\omega$ but $\epsilon = 1$
we find again a sign of $\gamma$ opposite to that of CHT06.

\subsubsection*{Time--averaged mean fields}

\begin{figure}\begin{center}
\includegraphics[width=\columnwidth]{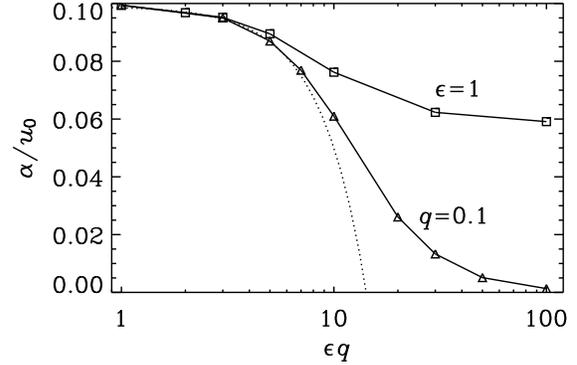}
\end{center}\caption[]{
Dependence of $\alpha / u_0$ on $\epsilon q$ for $\Rm=0.1$,
with $\epsilon=1$ and with $q = 0.1$.
The dotted curve corresponds to \eq{eq75}, which
has been derived for $\epsilon q \ll 1$ and $q \ll 1$ only.
}\label{palp_vs_eps}\end{figure}

Switching now to time--averaged mean fields we start with
\Fig{palp_vs_eps}, which
shows results for $\alpha$ at $\Rm=0.1$ in dependence on $\epsilon q$.
They were found with the help of numerical integrations of
the test--field version of \eq{eq11} in its complete form
or after reducing it to SOCA.
It turned out that SOCA is sufficient for their calculation.
Some of these results were also confirmed by evaluating \eq{eq43} with \eq{eq36} or \eq{eq45}.
As long as $\epsilon q$ is small, $\alpha$ depends in agreement with \eq{eq43} and \eq{eq45}
only via this product on $\epsilon$ and $q$.
For larger $\epsilon q$ it depends, however, in a more complex way on $\epsilon$ and $q$.
Furthermore, $\alpha$ remains finite if $\epsilon = 1$ and $q$ grows,
and it tends to zero if $q = 0.1$ and $\epsilon$ grows.
Since $\Rm$ is small the validity of SOCA is plausible in the case $q = 0.1$.
It is however remarkable in that with $\epsilon = 1$, in which $q$ may grow up to $10$.

\begin{figure}\begin{center}
\includegraphics[width=\columnwidth]{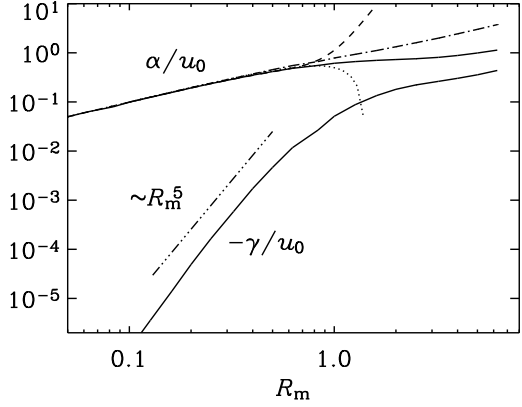}
\end{center}\caption[]{
Dependence of $\alpha/u_0$ and $-\gamma/u_0$ on $\Rm$ for $\epsilon=1$
and $\tilde\omega=1$ (i.e.\ $q = \Rm$).
Solid lines show results obtained without any approximation.
The dash-dotted line gives $\alpha/u_0$ as obtained numerically using SOCA.
The dotted and the dashed line give results calculated with \eq{eq43} and \eq{eq45},
or \eq{eq75}, respectively.
}\label{pscan_Rm}\end{figure}

Next, we consider the dependence of $\alpha$ and $\gamma$ on $\Rm$,
in \Fig{pscan_Rm} shown for $\epsilon=1$
and $\tilde{\omega} = 1$ (i.e. $q=\Rm$).
For small $\Rm$
we expect that SOCA applies and so $\alpha / u_0$ is linear in $\Rm$ but $\gamma$ vanishes.
Indeed $\alpha / u_0$ shows this linearity up to $\Rm \approx 1$.
In agreement with the results of CHT06  $\gamma$ is negative and its modulus remains small for $\Rm < 1$.
Remarkably the values of $\alpha / u_0$ calculated from \eq{eq43} and \eq{eq45} (dotted line),
or \eq{eq75} (dashed line),
which have been derived for $q \ll 1$ and $\epsilon q \ll 1$, deviate for $\Rm > 1$ drastically
from both the numerically obtained SOCA results (dash--dotted line)
and those obtained without any approximation of that kind (solid line).
The proportionality of $\gamma / u_0$ with $\Rm^5$ confirms the presumption made at the end of \Sec{HOCA}
that nonzero values of $\gamma$ occur only in sixth--order and higher approximations with respect to $u_0$.

\begin{figure}\begin{center}
\includegraphics[width=\columnwidth]{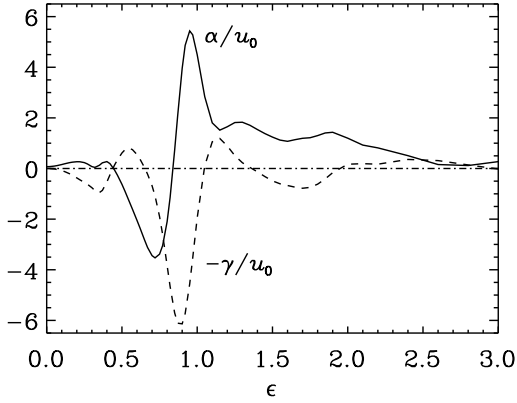}
\end{center}\caption[]{
Dependence of $\alpha/u_0$ and $-\gamma/u_0$ on $\epsilon$ for $\Rm=100$,
and $\tilde\omega=\sqrt{2/3}$ (the value considered by CHT06).
}\label{pscan_epsilon_om082}\end{figure}

\begin{figure}\begin{center}
\includegraphics[width=\columnwidth]{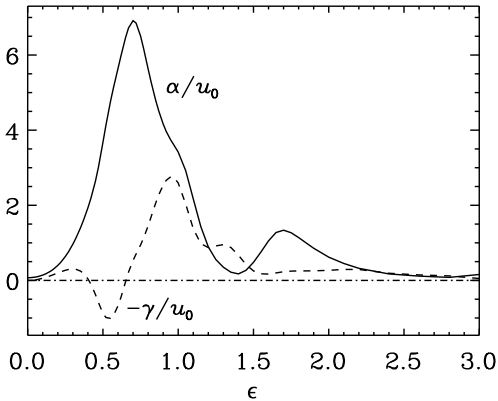}
\end{center}\caption[]{
Dependence of $\alpha/u_0$ and $-\gamma/u_0$ on $\epsilon$ for $\Rm=100$,
and $\tilde\omega=1$.
}\label{pscan_epsilon}\end{figure}

Simple arguments (as given in \Sec{Disc} below) suggest that $\alpha$ is never negative.
However, CHT06 found that not only the moduli but also
the signs of both $\alpha$ and $\gamma$ depend
for each given $\Rm$ sensitively on $\epsilon$.
In our \Fig{pscan_epsilon_om082}, which applies for $\Rm = 100$ and $\tilde{\omega} = \sqrt{2/3}$,
both $\alpha$ and $\gamma$ vary strongly with $\epsilon$, too.
The represented results confirm, apart from the sign of $\gamma$,
the corresponding ones in Fig.~2 of CHT06.
Both $\alpha$ and $\gamma$ change their signs with $\epsilon$.
As \Fig{pscan_epsilon} shows, in the situation with the same $\Rm$
and $\tilde{\omega} = 1$ only $\gamma$ changes its sign,
which indicates a considerable effect of changing $\tilde\omega$.
In both of the cases considered in \Fig{pscan_epsilon_om082} and \Fig{pscan_epsilon},
$\alpha$ and $\gamma$ diminish for small as well as large values of $\epsilon$.

\begin{figure}\begin{center}
\includegraphics[width=\columnwidth]{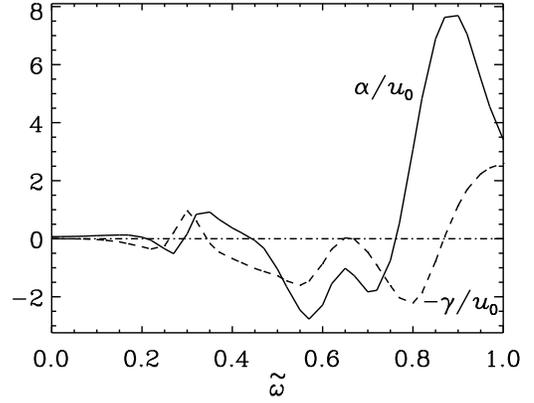}
\end{center}\caption[]{
Dependence of $\alpha/u_0$ and $-\gamma/u_0$ on $\tilde\omega$ for $\Rm=100$
and $\epsilon=1$.
}\label{pscan_omega}\end{figure}

\begin{figure}\begin{center}
\includegraphics[width=\columnwidth]{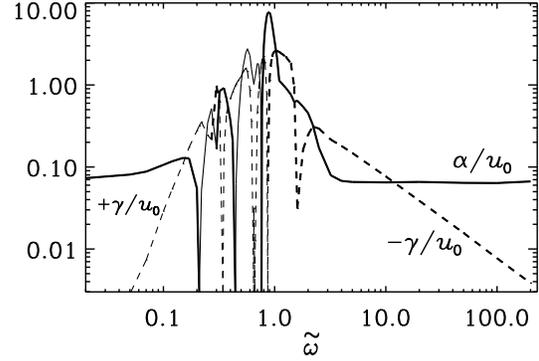}
\end{center}\caption[]{
Logarithmic representation of the dependence of
$+\alpha/u_0$ (thick solid lines), $-\alpha/u_0$ (thin solid lines),
$-\gamma/u_0$ (thick dashed lines), and $+\gamma/u_0$ (thin dashed lines)
on $\tilde\omega$ for $\Rm=100$ and $\epsilon=1$.
For large values of $\tilde\omega$ we have $\alpha/u_0\approx0.065$ and
$-\gamma/u_0\approx0.8/\tilde\omega$.
}\label{pscan_omega2}\end{figure}

In \Fig{pscan_omega} and \Fig{pscan_omega2} we see that $\alpha$ and $\gamma$
depend, at least for $\Rm =100$ and $\epsilon = 1$,
also sensitively on the parameter $\tilde\omega$, or $q$,
that is, on the frequency with which the velocity pattern wobbles.
There are, however, simple asymptotic behaviors for small and for large $\tilde{\omega}$,
clearly visible for $\tilde{\omega} < 0.1$ and $\tilde{\omega} > 3$.
Similar results have been found for $\Rm=10$ and $\epsilon = 1$.
In this case, however, $\alpha$ stays positive
for all values of $\tilde\omega$, and only one sign reversal of $\gamma$ occurs.

\begin{figure}\begin{center}
\includegraphics[width=\columnwidth]{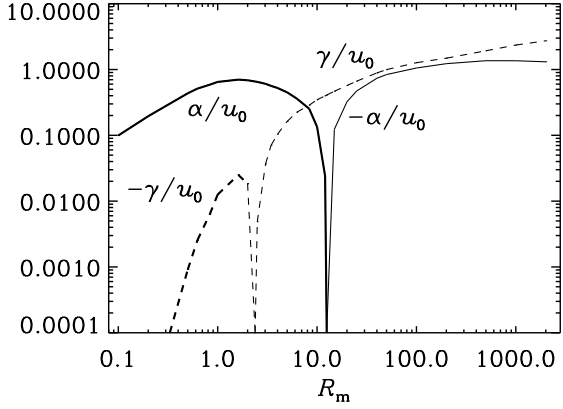}
\end{center}\caption[]{
Logarithmic representation of the dependence of $\alpha$ (thick solid line),
$-\alpha$ (thin solid line), $-\gamma$ (thick dashed line),
and $+\gamma$ (thin dashed line) on $\Rm$
for $\tilde\omega=0.5$ for $\epsilon=1$.
}\label{pscan_Rm_om05}\end{figure}

\begin{figure}\begin{center}
\includegraphics[width=\columnwidth]{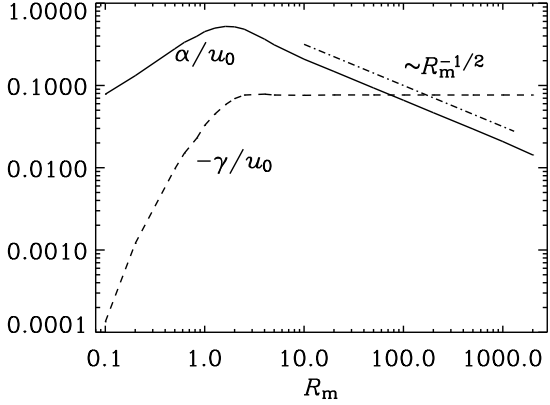}
\end{center}\caption[]{
Dependence of $\alpha$ and $\gamma$ on $\Rm$
for $\tilde\omega=10$ for $\epsilon=1$.
Note the asymptotic scaling
$\alpha/u_0\sim\Rm^{-1/2}$ (dash--dotted line)
and that $\gamma$ approaches a constant finite value as $\Rm \to \infty$.
}\label{pscan_Rm_om10}\end{figure}

We see from CHT06 that there is
a rich dependence of $\alpha$ and $\gamma$ on $\Rm$
for values of $\tilde\omega$ and $\epsilon$ of order unity.
In \Fig{pscan_Rm_om05} we show
results for an example with $\tilde\omega=0.5$.
Reversals of $\alpha$ are then possible for rather small values of
$\Rm$ of the order of 10.
However, as \Fig{pscan_Rm_om10} shows, such behaviour disappears for
$\tilde\omega=10$, in which case $\alpha$
stays always positive and $\gamma$ always negative.
In fact, there is an asymptotic scaling $\alpha/u_0\sim\Rm^{-1/2}$
as $\Rm \to \infty$,
and $\gamma$ approaches a constant finite value as $\Rm \to \infty$.

\begin{figure}\begin{center}
\includegraphics[width=\columnwidth]{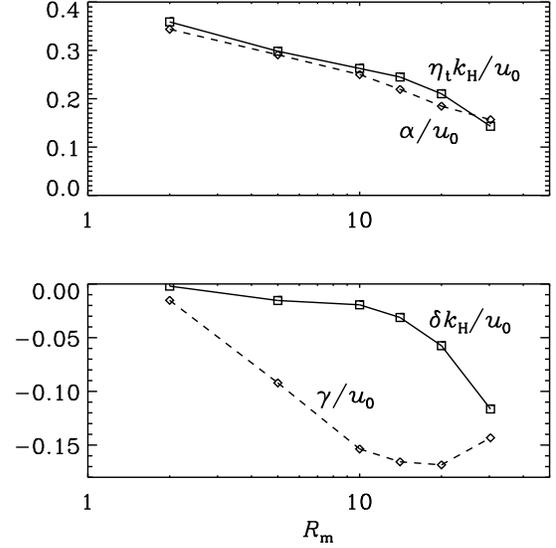}
\end{center}\caption[]{
Dependence of $\alpha$ and $\gamma$ (dashed lines) as well as
$\eta_{\rm t}$ and $\delta$ (solid lines) on $\Rm$,
for $\tilde\omega=0.7$, $\epsilon=1$, and $k=k_{\rm H}$.
}\label{p3D_RmDep_eps1_om07}\end{figure}

In a few cases $\eta_{\rm t}$ and $\delta$ have been determined
in addition to $\alpha$ and $\gamma$.
Results on the dependence of these quantities with $\epsilon = 1$ and $\tilde{\omega} = 0.7$ on $\Rm$
are shown in \Fig{p3D_RmDep_eps1_om07}.
They have however been calculated with $k = k_{\rm H}$, not $k \to 0$, and are therefore
at most approximations of the mentioned quantities.

A correct interpretation of these results requires a look on the explanations of \Sec{Generalization}
on the non-local connection between $\meanEMF$, $\meanBB$ and $\meanJJ$
as defined by \eq{eqxx01}.
In that sense the $\alpha$, $\gamma$, $\eta_{\rm t}$ and $\delta$ in \Fig{p3D_RmDep_eps1_om07}
may be understood as values of the functions $\tilde{\alpha} (k)$, $\tilde{\gamma} (k)$,
$\tilde{\eta}_{\rm t} (k)$ and $\tilde{\delta} (k)$ at $k = k_{\rm H}$.

In the following, when writing $\alpha (k)$ or $\eta_{\rm t} (k)$, for example,
we always mean $\tilde{\alpha} (k)$ or $\tilde{\eta}_{\rm t} (k)$.
In an earlier investigation with the Roberts flow under SOCA
and with isotropic turbulence independent of SOCA \citep{Betal08b} it was found that
$\alpha (k)$ and $\eta_{\rm t} (k)$ vary with $k$
in a Lorentzian fashion like $(1+k^2/k_{\rm H}^2)^{-1}$.
However, for the GP flow \cite{Cou08} found that
$\alpha(k)$ at small $k$ is extremely sensitive to the value of $\Rm$.

\begin{figure}\begin{center}
\includegraphics[width=\columnwidth]{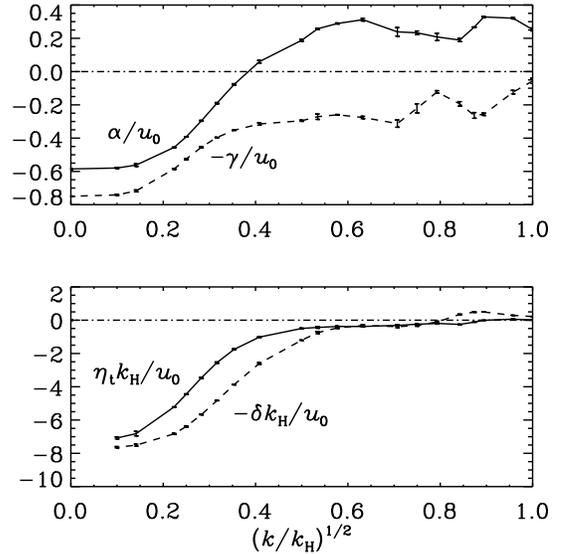}
\end{center}\caption[]{
Functions $\alpha (k)$, $\gamma (k)$, $\eta_{\rm t} (k)$, and $\delta (k)$
for $\Rm=30$, $\tilde\omega=0.5$, and $\epsilon=1$.
}\label{pscan_k_Rm30}\end{figure}

\Fig{pscan_k_Rm30} shows that $\alpha(k)$ and $\gamma(k)$ for
$\Rm=30$, $\tilde\omega=0.5$, and $\epsilon=1$, approach the values
given in \Fig{pscan_Rm_om05} as $k \to 0$.
However, the magnitudes of $\eta_{\rm t}(k)$ and $\delta(k)$ become
rather large as $k\to 0$.
It turns out that $\alpha$ is positive for $k/k_{\rm H}>0.4$ and
$\gamma$ becomes smaller with increasing $k$.
Remarkably, $\eta_{\rm t}(k)$ is negative for $k/k_{\rm H}<1$,
suggesting that magnetic field generation might be possible
via a negative magnetic diffusion instability.

\begin{figure}\begin{center}
\includegraphics[width=\columnwidth]{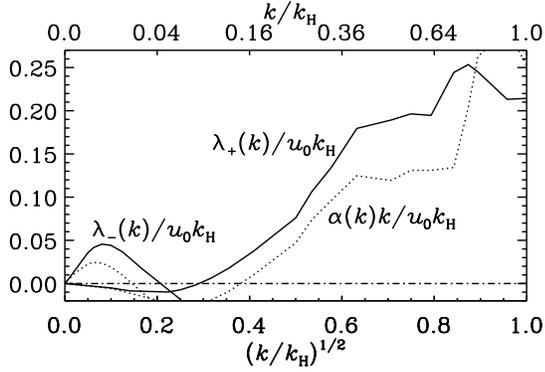}
\end{center}\caption[]{
Dependence of $\lambda_\pm$ on $k$ for $\Rm=30$, $\tilde\omega=0.5$,
and $\epsilon=1$.
The dotted lines give $\pm\alpha k$ for comparison.
}\label{pscan_k_Rm30_lam}\end{figure}

In order to check this possibility
we have calculated the linear growth rates
\EQ
\lambda_\pm(k)=-[\eta+\eta_{\rm t} (k)]k^2 \pm\alpha (k) k \, .
\EN
\Fig{pscan_k_Rm30_lam} shows that $\lambda_\pm$
is almost entirely given by $\pm\alpha(k)k$.
A negative diffusivity instability does not occur.
It is important to realize that most of the small wavenumber modes,
especially those with negative values of $\eta_{\rm t}$,
would never be realized.
This is because in a system of given size, only the corresponding
harmonics will have a chance to be excited, and of those only the
ones with the largest growth rates will dominate.
We should point out that the detailed variations of $\lambda_+$ shown
in \Fig{pscan_k_Rm30_lam} may not be accurate.
In fact, this figure shows a maximum at $k/k_{\rm H}\approx0.75$,
but direct simulations suggest that the fastest growth occurs
for $k/k_{\rm H}\approx0.5$ with a grow rate of
$\lambda\approx0.23 u_{\rm rms}k_{\rm f}$.
Nevertheless, this value is still compatible with \Fig{pscan_k_Rm30_lam}.

\begin{figure}\begin{center}
\includegraphics[width=\columnwidth]{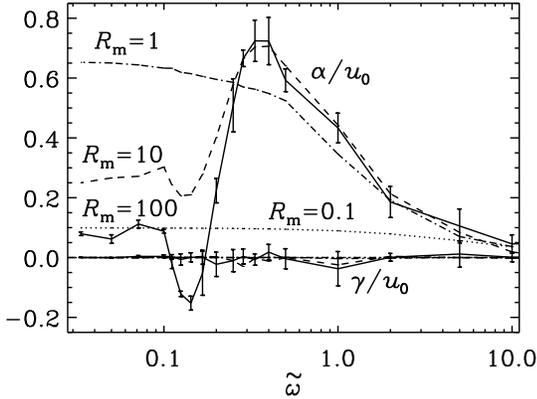}
\end{center}\caption[]{
Dependence of $\alpha$ and $\gamma$ on $\tilde\omega$
for the flow with random time dependence (case ii),
with $\Rm=100$ (solid lines), $\Rm=10$ (dashed lines),
$\Rm=1$ (dash-dotted lines), and $\Rm=0.1$ (dotted lines),
and with $\epsilon = \pi$.
The error bars are similar in all cases, but are only shown for $\Rm=100$.
}\label{pscan_tau_RandomPhase}\end{figure}

\subsection{Case (ii)}

In case (ii) we have calculated $\alpha$ and $\gamma$
under the assumptions on $\phi_x$ and $\phi_y$ introduced in \Sec{Randomflow}.
\Fig{pscan_tau_RandomPhase} shows results for $\Rm$ ranging from 0.1 to 100 and $\epsilon = \pi$
as functions of $\tilde\omega$.
In the limit of small $\tilde\omega$ the flow can be considered as stationary,
that is, as a Roberts flow.
Indeed in this limit the values of $\alpha$ agree well with those obtained
for the Roberts flow
(e.g., \cite{Radler_etal02a}, see also \App{ApRob}).
For large values of $\tilde\omega$ the values of $\alpha$ vanish for all $\Rm$.
For not too small $\tilde\omega$ and $\Rm$ there is no longer a noticeable variation of $\alpha$
with $\Rm$, and $\alpha$ reaches a maximum at $\tilde\omega\approx0.3$.
In the range $0.3\leq\tilde\omega\leq1$ continuous flow renewal removes
the tendency for $\alpha$ to diminish with growing $\Rm$.
For $\tilde\omega\leq0.2$ the value of $\alpha$ remains strongly
dependent on $\Rm$ and can still change sign.
We have also calculated $\alpha$ with $\epsilon=1$ and $\Rm=100$
as a function of $\tilde\omega$ and found a
qualitatively similar behavior as for $\epsilon=\pi$.
In this case it remains positive and
is up to 50\% smaller than for $\epsilon=\pi$ when $\tilde\omega<1$
and somewhat larger when $\tilde\omega>1$.
In all cases we found, as expected, $\gamma=0$ within error margins.

\section{Discussion}
\label{Disc}

Our results for the flow of type (i)
confirm the finding of CHT06 that both the $\alpha$ and $\gamma$ coefficients
depend sensitively on $\Rm$ and also on $\epsilon$,
and that even the signs of these coefficients may vary with these parameters.
We have to add that $\alpha$ and $\gamma$ depend also on $\tilde\omega$, or $q = \tilde\omega \, \Rm$,
that is, on parameters connected with the frequency of the wobbling motion,
which CHT06 fixed in a special way without commenting on it,
and that they show similar variations with these parameters.
We found however rather regular behaviors of $\alpha$ and $\gamma$
for small and for large values of $\epsilon$ and $\tilde\omega$.

It is sometimes considered as a rule that the sign of $\alpha$
is opposite to that the mean helicity of the fluid flow and its modulus
is proportional to that of the mean helicity.
There is however no general reason for that kind of relation between $\alpha$ and the kinetic helicity.
We see only two limiting cases which allow simple statements on the sign of $\alpha$.

Firstly, in the framework of SOCA applied to
homogeneous turbulence and comparable flows it turns out
that the sign of $\alpha$ in the limit $q \to 0$ is always opposite
to that of $\ol{\ppsi \cdot (\nab \x \ppsi)}$,
where $\uu = \nab \x \ppsi$, $\nab \cdot \ppsi = 0$;
see, e.g., \cite{KR80,RB03}.
For both types of flows, (i) and (ii),
we have $\ppsi = (\tilde\psi, - \tilde\psi, \psi)$,
where $\tilde\psi = - k_{\rm H} (\p \psi/\p x + \p \psi/\p y)$
and therefore $\ol{\ppsi \cdot (\nab \x \ppsi)} = - 2 u_0^2 / k_{\rm H}$.
This implies that $\alpha$ is positive.
Indeed,
only positive $\alpha$ have been observed for small $q$, even beyond SOCA.

Secondly it was found in SOCA under the same conditions, but in the limit $q \to \infty$,
that the sign of $\alpha$ for a flow with finite correlation time is opposite to that of
$\int_0^\infty \ol{\uu (\xx, t) \cdot (\nab \x \uu (\xx, t - \tau)} \, \dd \tau$;
see, e.g., again \cite{KR80}.
A relation of that kind between $\alpha$ and this integral can indeed be formally derived
from the general relation (29) of \cite{RR07} and applied to our specific situation.
In case (i) the correlation time is however infinite and this integral does not converge.
Although we know that $\ol{\uu \cdot (\nab \x \uu)} = - 2 u_0^2 k_{\rm H}$ we do not see
how reliable conclusions could be drawn concerning the sign of $\alpha$ in the limit $q \to \infty$.
In case (ii)
the integral is positive, and indeed only positive $\alpha$ have been observed.

Beyond the low and high conductivity limits, that is, for not too small or not too large values of $q$,
even SOCA offers no simple general statements on the sign of $\alpha$.
In general $\alpha$ may take both positive and negative values.

For studying $\alpha$ and $\eta_{\rm t}$ with very simple flows it seems appropriate
to consider flows of type (ii) rather than of type (i).
In case (i) the results are influenced by the aforementioned circular motion of the flow pattern.
As long as only $\alpha$ and $\eta_{\rm t}$ should be discussed
there is hardly a reason to introduce such a motion.
We see no natural interpretation of it and so no interpretation
of the so caused $\gamma$ and $\delta$ effects.

Recently \cite{Tilgner08} pointed out that a time--dependent flow
of a conducting fluid can act as a dynamo even when steady flows which coincide
with it at any particular time cannot.
He demonstrated this with a Roberts flow modified by a drift of its pattern
so that the velocity $\uu$ satisfies relations like \eq{eq03} and \eq{eq05}
with $\varphi_x = - k_{\rm H} v_{\rm d} t$ and $\varphi_y = 0$,
where $v_{\rm d}$ is constant the drift velocity.
Even if the intensity of the flow is too weak so that in the case $v_{\rm d} = 0$
no growing solutions of the induction equation with a given period in the $z$ direction exist,
such solutions may occur in an interval of some finite $v_{\rm d}$.
Although this flow considered by Tilgner is in a sense simpler than the flows
in our paper, it shows no longer the symmetries with respect to the $z$ axis
which we have utilized.
As a consequence the relation between the mean electromotive force $\meanEMF$
and the mean magnetic field $\meanBB$ is more complex.
In particular \eq{eq22} and \eq{eq23} no longer apply.
Nevertheless the question arises whether the effect of the time--dependence of flows
observed by Tilgner occurs also in the examples investigated here.
In case (i) the parameter $\omega$ could play the role of $v_{\rm d}$.
The fact that the magnitude of $\alpha$ is larger for some finite $\tilde{\omega}$
than for $\tilde{\omega} = 0$, which can be seen in \Figs{pscan_omega}{pscan_omega2},
points in this direction.

One of the original motivations for looking at the GP flow was the fact
that it is time-dependent and in that sense closer to turbulent flows than
time-independent flows.
However, as we have shown here, the dynamo properties of the GP flow
cannot be compared in a meaningful way with analytic theories or with
simulations that apply to isotropic turbulence.
Nevertheless, as shown in this paper, an analytic theory for the
$\alpha$ effect and other turbulent transport coefficients can be
derived that matches numerical results in limiting cases.

Astrophysical flows can often neither be described by isotropic turbulence
nor by wobbling two-dimensional flow patterns, but they are likely to
contain aspects of both extremes.
However, the present work highlights another aspect that may
be of more general signi\-fi\-cance and concerns the turbulent transport
properties in the presence of high-frequency time variability.
This is not just a peripheral aspect of turbulence, but it is an
additional property whose effects need to be understood more thoroughly.
The situation is reminiscent of the modifications of mixing length theory
in the presence of stellar pulsations \citep[see, e.g.,][]{Gou77}.
In dynamo theory the issue of high-frequency time variability
has only recently been addressed.
One example concerns the nonlinear $\alpha$ effect where its
time dependence has a striking effect on the behaviour of the mean field.
In that example the temporal behaviour of the forcing function
(delta-correlated or steady) determines the nonlinear asymptotic scaling
behavior of the quenching function $\alpha(\meanBB)$ at low $\Rm$.
The early results of \cite{Mof72} and \cite{Rue74} suggested a
$|\alpha|\sim|\meanBB|^{-3}$ behavior, but in more recent years
\cite{FBC99} and \cite{RK00} found instead a $|\alpha|\sim|\meanBB|^{-2}$
behavior, which seemed in conflict with the earlier results.
However, the work of \cite{Sur_etal07} now shows that this is
not just an artifact related to different approximations, for example,
but it depends on whether or not the flow is time-dependent.
They found that the $|\alpha|\sim|\meanBB|^{-3}$ behavior is
reproduced if the flow is steady, while the $|\alpha|\sim|\meanBB|^{-2}$
behavior is obtained in the time-dependent case using a forcing function
that is $\delta$-correlated in time.
Again, it is not clear which types of flows are more astrophysically
relevant, but it is now clear that the detailed time-dependence of the
turbulent flows can affect its transport properties in rather unexpected
ways.

\section*{Acknowledgments}

We acknowledge Nordita and the Kavli Institute for Theoretical Physics
for providing a stimulating atmosphere during their programs on dynamo theory
in 2008.
This research was supported in part by the National Science
Foundation under grant PHY05-51164.

\newcommand{\ybook}[3]{ #1, {#2} (#3)}
\newcommand{\yjfm}[3]{ #1, {J.\ Fluid Mech.,} {#2}, #3}
\newcommand{\yprl}[3]{ #1, {Phys.\ Rev.\ Lett.,} {#2}, #3}
\newcommand{\ypre}[3]{ #1, {Phys.\ Rev.\ E,} {#2}, #3}
\newcommand{\yapj}[3]{ #1, {ApJ,} {#2}, #3}
\newcommand{\ynat}[3]{ #1, {Nat,} {#2}, #3}
\newcommand{\yptrsa}[3]{ #1, {Phil. Trans. Roy. Soc. London A,} {#2}, #3}
\newcommand{\ymn}[3]{ #1, {MNRAS,} {#2}, #3}
\newcommand{\yan}[3]{ #1, {AN,} {#2}, #3}
\newcommand{\yana}[3]{ #1, {A\&A,} {#2}, #3}
\newcommand{\ygafd}[3]{ #1, {Geophys.\ Astrophys.\ Fluid Dyn.,} {#2}, #3}
\newcommand{\ypf}[3]{ #1, {Phys.\ Fluids,} {#2}, #3}
\newcommand{\yproc}[5]{ #1, in {#3}, ed.\ #4 (#5), #2}
\newcommand{\yjour}[4]{ #1, {#2} {#3}, #4.}
\newcommand{\sapj}[1]{ #1, {ApJ} (submitted)}
\newcommand{\pmn}[1]{ #1, {MNRAS} (in press)}


\label{lastpage}

\appendix

\section{Roberts flow}
\label{ApRob}

In the special case $\epsilon = 0$
the Galloway--Proctor flow, defined by \eq{eq03} and \eq{eq05}, turns
into the Roberts flow.
Our SOCA results for this special case agree with results for the Roberts flow reported
in Brandenburg et al.\ (2008) and in R\"adler et al.\ (2002a), referred to as BRS08 and R02a, respectively.

In BRS08 instead of our coordinate system $(x, y, z)$ another one, say $(x', y', z)$, is used,
which is obtained by a $45^\circ$ rotation of our system about the $z$ axis, that is,
\EQ
x = \frac{1}{\sqrt{2}} (x' - y') \, , \quad y = \frac{1}{\sqrt{2}} (x' + y') \, .
\label{c01}
\EN
In the case $\epsilon = 0$, to which we restrict ourselves here,
\eq{eq05} turns under this transformation into
\EQ
\psi = \frac{2 u_0}{k_{\rm H}} \cos \left(k_{\rm H} x' / \sqrt{2}\right) \, \cos \left(k_{\rm H}x'/\sqrt{2}\right) \, .
\label{c03}
\EN
Together with \eq{eq03} we find so, referring to the system $(x', y', z)$,
\EQ
\uu=\sqrt{2} u_0\pmatrix{
-\cos \left(k_{\rm H}x'/\sqrt{2}\right) \,
 \sin \left(k_{\rm H}y'/\sqrt{2}\right) \cr
+\sin \left(k_{\rm H}x'/\sqrt{2}\right) \,
 \cos \left(k_{\rm H}y'/\sqrt{2}\right) \cr
 -\sqrt{2} \cos \left(k_{\rm H}x'/\sqrt{2}\right) \,
           \cos \left(k_{\rm H}y'/\sqrt{2}\right) } \, .
\label{c05}
\EN

Comparing this first with (BRS08 25) and ignoring the opposite sign of $u_z$ we find
\EQ
u_0^{\rm BRS} = \sqrt{2} \, u_0 \, , \quad k_0 = k_{\rm f} / \sqrt{2} = k_{\rm H} / \sqrt{2} \, .
\label{c07}
\EN
The only consequence of inverting the sign of $u_z$ is a sign change of $\alpha$.
Taking then the SOCA results (BRS08 30) for $\alpha$ and $\eta_{\rm t}$,
with $u_0^{\rm BRS}$ in place of $u_0$ and completed by (BRS08 29),
considering \eq{c07} and the remark on the sign of $\alpha$ we can easily reproduce
our results \eq{eq43}.

This applies analogously to R02a if,
in addition to the transformation \eq{c01}, $x'$
is replaced by $x' - \pi / \sqrt{2} k_{\rm H}$ and $y'$ by $y' + \pi / \sqrt{2} k_{\rm H}$.
Comparing the corresponding modification of \eq{c05} with (R02a 15) we find
\EQ
u_\perp = (2 \sqrt{2} / \pi) u_0 \, , \quad u_\parallel = (8 / \pi^2) u_0 \, , \quad
    a = \sqrt{2} \pi / k_{\rm H} \, .
\label{c09}
\EN
When using (R02a 19) we obtain our result \eq{eq43} for $\alpha$.
With (R02a 38) and $\eta_{\rm t} = \beta_\perp + \beta_3$ we may also reproduce
our result \eq{eq43} for $\eta_{\rm t}$.

Going beyond SOCA we note that according to (BRS08 25), or also according to (R02a 20),
\EQ
\alpha = u_0 \Rm \phi (2 \Rm)
\label{eqc11}
\EN
with a function $\phi$ satisfying $\phi (0) = 1$ and vanishing like $\Rm^{- 3/2}$ with growing $\Rm$.
It has been calculated numerically and is plotted, e.g., in R02a.

\section{Second--order calculations}
\label{ApSOCA}

For the calculation of the $\alpha_{ij}$ and $\eta_{ij}$ with $1 \leq i, j \leq 2$
under SOCA we start with \eq{eq31}.
Introducing there $\bb = \Rey (\hat{\bb} \exp \ii k z)$
and $\meanBB = \Rey (\hat{\BB} \exp \ii k z)$
we obtain
\EQ
\left[\p_t - \eta(\nab^2 - k^2)\right] \hat{\bb} = (\hat{\BB} \cdot \nab) \uu - \ii k u_z \hat{\BB} \, .
\label{a01}
\EN
For our purposes it is useful to represent $\uu$ in the form
\EQA
u_x \!\!\! &=& \!\!\! - u_0 \big[ \cos (k_{\rm H} y) \, \mbox{ss} (t) + \sin (k_{\rm H} y) \, \mbox{cs} (t) \big] \, ,
\nonumber\\
u_y \!\!\! &=& \!\!\! u_0 \big[ \cos (k_{\rm H} x) \, \mbox{sc} (t) + \sin (k_{\rm H} x) \, \mbox{cc} (t) \big] \, ,
\nonumber\\
u_z \!\!\! &=& \!\!\! - u_0 \big[ \cos (k_{\rm H} x) \, \mbox{cc} (t) - \sin (k_{\rm H} x) \, \mbox{sc} (t)
\label{a03}\\
&&    + \cos (k_{\rm H} y) \, \mbox{cs} (t) - \sin (k_{\rm H} y) \, \mbox{ss} (t) \big] \, ,
\nonumber
\ENA
where
\EQ
\mbox{ss} (t) = \sin (\epsilon \sin \omega t) \, , \quad
\mbox{cs} (t) = \cos (\epsilon \sin \omega t) \, , \quad \mbox{etc.}
\label{a05}
\EN
Then the right--hand side of \eq{a01}, say $\hat{\RR}$, takes then the form
\EQA
\hat{\RR} &=& \hat{\RR}^{{\rm c} x} \cos k_{\rm H} x + \hat{\RR}^{{\rm s} x} \sin k_{\rm H} x
\nonumber\\
&&  + \hat{\RR}^{{\rm c} y} \cos k_{\rm H} y + \hat{\RR}^{{\rm s} y} \sin k_{\rm H} y
\label{a07}
\ENA
with $\hat{\RR}^{{\rm c} x}$, $\hat{\RR}^{{\rm s} x}$, $\cdots$ depending on time.

Clearly, \eq{a01} poses an initial value problem.
As initial time $t_0$ we take $t_0 \to - \infty$.
Then the solution $\hat{\bb}$ of \eq{a01} is completely determined by its right--hand side, $\hat{\RR}$,
and has again the form of $\hat{\RR}$ as given by \eq{a07}.
Since $\nab^2 \hat{\bb} = - k_{\rm H}^2 \hat{\bb}$ we have
\EQ
\hat{\bb} (t) = \int_0^\infty \!\!\!\! \hat{\RR} (t - t') \exp [- \eta (k_{\rm H}^2 + k^2) t'] \, \dd t' \, .
\label{a11}
\EN
After determining the $\hat{\RR}^{{\rm c} x}$, $\hat{\RR}^{{\rm s} x}$, $\cdots$
we find for the analogously defined $\hat{\bb}^{{\rm c} x}$, $\hat{\bb}^{{\rm s} x}$, $\cdots$
\EQA
\hat{b}_x^{{\rm c} x} &=& \ii u_0 k \hat{B}_x \Gamma (\mbox{cc}) \, , \quad
    \hat{b}_x^{{\rm s} x} = - \ii u_0 k \hat{B}_x \Gamma (\mbox{sc}) \, ,
\nonumber\\
\hat{b}_x^{{\rm c} y} &=& - u_0 (k_{\rm H}  \hat{B}_y - \ii k \hat{B}_x) \Gamma (\mbox{cs}) \, ,
\nonumber\\
\hat{b}_x^{{\rm s} y} &=& u_0 (k_{\rm H}  \hat{B}_y - \ii k \hat{B}_x) \Gamma (\mbox{ss}) \, ,
\nonumber\\
\hat{b}_y^{{\rm c} x} &=& u_0 (k_{\rm H}  \hat{B}_y + \ii k \hat{B}_x) \Gamma (\mbox{cc}) \, ,
\nonumber\\
\hat{b}_y^{{\rm s} x} &=& - u_0 (k_{\rm H}  \hat{B}_y + \ii k \hat{B}_x) \Gamma (\mbox{sc}) \, ,
\label{a13}\\
\hat{b}_y^{{\rm c} y} &=& \ii u_0 k \hat{B}_y \Gamma (\mbox{cs}) \, , \quad
    \hat{b}_y^{{\rm s} y} = - \ii u_0 k \hat{B}_y \Gamma (\mbox{ss}) \, ,
\nonumber\\
\hat{b}_z^{{\rm c} x} &=& u_0 k_{\rm H} \hat{B}_x \Gamma (\mbox{sc}) \, , \quad
    \hat{b}_z^{{\rm s} x} = u_0 k_{\rm H} \hat{B}_x \Gamma (\mbox{cc}) \, ,
\nonumber\\
\hat{b}_z^{{\rm c} y} &=& u_0 k_{\rm H} \hat{B}_y \Gamma (\mbox{ss}) \, , \quad
    \hat{b}_z^{{\rm s} y} = u_0 k_{\rm H} \hat{B}_y \Gamma (\mbox{cs}) \, ,
\nonumber
\ENA
where
\EQ
\Gamma (f) = \int_0^\infty \!\!\!\! f (t - t') \exp [- \eta (k_{\rm H}^2 + k^2) t'] \, \dd t'
\label{a15}
\EN
with any function $f = f (t)$.
Of course, $\Gamma$ depends in general on time.
In view of \eq{a13} we note that, since $\meanBB = \Rey (\hat{\BB} \exp\ii k z)$,
we have also $\meanJJ = \Rey (\hat{\JJ} \exp\ii k z)$
and therefore $ik \hat{B}_x = \hat{J}_y$ and $ik \hat{B}_y = - \hat{J}_x$.

Calculating then $\meanEMF$ we find
\EQA
\meanemf_x \!\!\! &=& \!\!\! \ol{u_y b_z} - \ol{u_z b_y}
    = u_0 \Rm a_{xx} \meanB_x - u_0 \Rm b_{xx} \meanJ_x
\nonumber\\
\meanemf_y \!\!\! &=& \!\!\! \ol{u_z b_x} - \ol{u_x b_z}
    = u_0 \Rm a_{yy} \meanB_y  - u_0 \Rm b_{yy} \meanJ_y
\label{a17}\\
\meanemf_z \!\!\! &=& \!\!\! \ol{u_x b_y} - \ol{u_y b_x} = 0 \, .
\nonumber
\ENA
where $a_{xx}$, $a_{yy}$, $b_{xx}$ and $b_{yy}$ are in general periodic functions of time,
which are defined by
\EQA
a_{xx} \!\!\! &=& \!\!\! \eta k_{\rm H}^2 \left[ \mbox{cc} \Gamma (\mbox{cc})
    + \mbox{sc} \Gamma (\mbox{sc}) \right] \, ,
\nonumber\\
a_{yy} \!\!\! &=& \!\!\! \eta k_{\rm H}^2 \left[ \mbox{cs} \Gamma (\mbox{cs})
    + \mbox{ss} \Gamma (\mbox{ss}) \right] \, ,
\label{a19}\\
b_{xx} \!\!\! &=& \!\!\! b_{yy} = \half \left( a_{xx} + a_{yy} \right) \, .
\nonumber
\ENA
The combination of trigonometric functions on the right--hand side of the relation
for $a_{xx}$ can easily be expressed by the function $\mbox{CC}$ defined by \eq{eq37}.
The same applies to $a_{yy}$ and the function $\mbox{CS}$ defined in \eq{eq67}.

The result given by \eq{a17} and \eq{a19} is valid for arbitrary $k$.
This applies of course also if it is written in the alternative form with $\mbox{CC}$ and $\mbox{CS}$.
In that sense it is of some interest in view of the nonlocal connection between $\meanEMF$ and $\meanBB$
studied in the paper by \cite{Betal08b}.
In the main part of the present paper we consider however the limit $k \to 0$ only.
In this limit \eq{a15} applies with $k = 0$.
Then \eq{a17} and \eq{a19} agree just with \eq{eq33} and \eq{eq37}.

\begin{table}
\begin{center}
\caption{Some values of $\chi^{(2)}$ defined in \Eq{eq36}.
\label{tab1}}
\begin{tabular}{crrrrrr}
\hline
$\!\!{\displaystyle{\omega t\over\pi/8}}\!\!$ & $\epsilon=0.1\!$ & $\epsilon=0.1\!$ & $\epsilon=1$ & $\epsilon=1$ &
$\epsilon=10\!$ & $\epsilon=10\!$ \cr
& $q=1$ & $q=10$ & $q=1$ & $q=10\!\!$ & $q=1$ & $q=10\!$ \cr
\hline
$0\!$&$\! 0.9970\!$&$\! 0.9926\!$&$\! 0.7558\!$&$\! 0.4204\!$&$\! 0.3017\!$&$\! 0.2073\!$\\
$1\!$&$\! 0.9982\!$&$\! 0.9936\!$&$\! 0.8521\!$&$\! 0.4882\!$&$\! 0.5246\!$&$\! 0.2667\!$\\
$2\!$&$\! 0.9990\!$&$\! 0.9954\!$&$\! 0.9123\!$&$\! 0.6165\!$&$\!-0.1882\!$&$\!-0.1816\!$\\
$3\!$&$\! 0.9989\!$&$\! 0.9971\!$&$\! 0.8963\!$&$\! 0.7341\!$&$\! 0.1567\!$&$\! 0.1912\!$\\
$4\!$&$\! 0.9980\!$&$\! 0.9975\!$&$\! 0.8115\!$&$\! 0.7658\!$&$\!-0.1558\!$&$\!-0.2448\!$\\
$5\!$&$\! 0.9968\!$&$\! 0.9965\!$&$\! 0.7121\!$&$\! 0.6889\!$&$\! 0.1196\!$&$\! 0.1881\!$\\
$6\!$&$\! 0.9960\!$&$\! 0.9947\!$&$\! 0.6586\!$&$\! 0.5550\!$&$\!-0.0363\!$&$\!-0.1637\!$\\
$7\!$&$\! 0.9961\!$&$\! 0.9931\!$&$\! 0.6775\!$&$\! 0.4465\!$&$\! 0.1317\!$&$\! 0.2189\!$\\
$8\!$&$\! 0.9970\!$&$\! 0.9926\!$&$\! 0.7558\!$&$\! 0.4204\!$&$\! 0.3017\!$&$\! 0.2073\!$\\
\hline
\end{tabular}
\end{center}
\end{table}

\section{Higher--order calculations}
\label{ApHOCA}

For the sake of simplicity we assume now, beyond SOCA, that $\meanBB$ is a uniform field,
that is, has no spatial derivatives.
Then $\ol{\uu \x \bb}$ is independent of space coordinates and \eq{eq59} turns into
\EQ
(\p_t - \eta \nab^2) \bb^{(n+1)} = (\bb^{(n)} \cdot \nab) \uu - (\uu \cdot \nab) \bb^{(n)} \, , \quad  n \geq 1 \, .
\label{b01}
\EN
We may apply some modification of the procedure used in \App{ApSOCA} for solving the equation \eq{a01}
for $\hat{\bb}$ to the equations \eq{b01} for $\bb^{(2)}$ and $\bb^{(3)}$.

The right--hand side of the equation for $\bb^{(2)}$, say $\RR^{(2)}$,
is a linear combination of products $\varphi (x) \, \varphi (y)$,
where $\varphi (x)$ stands for $\cos k_{\rm H} x$ or $\sin k_{\rm H} x$,
and $\varphi (y)$ for $\cos k_{\rm H} y$ or $\sin k_{\rm H} y$.
Clearly $\bb^{(2)}$ has the same form as $\RR^{(2)}$ and satisfies $\nab^2\bb^{(2)} = - 2 k_{\rm H}^2 \bb^{(2)}$.
Therefore \eq{a11} applies after replacing $\hat{\bb}$ and $\hat{\RR}$ by $\bb^{(2)}$ and $\RR^{(2)}$, respectively,
$k_{\rm H}^2$ by $2 k_{\rm H}^2$, and putting $k = 0$.
As a consequence of the described structure of $\bb^{(2)}$ we have $\ol{\uu \x \bb^{(2)}} = \bzo$.

The right--hand side of the equation for $\bb^{(3)}$, which we call $\RR^{(3)}$, is a linear combination
of products $\varphi_1 (x) \, \varphi_2 (x) \, \varphi (y)$ or $\varphi (x) \, \varphi_1 (y) \, \varphi_2 (y)$,
where the indices $1$ and $2$ may refer to the same function or to different functions,
e.g., $\varphi_1 (x) = \varphi_2 (x) = \cos k_{\rm H} x$,
or $\varphi_1 (x) = \cos k_{\rm H} x$ and $\varphi_2 (x) = \sin k_{\rm H} x$.
In the first case we utilize $\cos^2 k_{\rm H} x = \half (1 + \cos 2 k_{\rm H} x)$
and split, e.g., $\cos^2 k_{\rm H} x \, \sin k_{\rm H} y$ into the two parts $\half \sin k_{\rm H} y$
and $\half \cos 2 k_{\rm H} x \, \sin k_{\rm H} y$.
In this way we may split $\RR^{(3)}$ into two parts, $\RR^{(3 a)}$ and $\RR^{(3b)}$,
where $\RR^{(3 a)}$ contains only contributions $\varphi_1 (x) \, \varphi_2 (x) \varphi (y)$
and $\varphi (x) \, \varphi_1 (y) \varphi_2 (y)$ with three different factors,
and contributions of the types $\varphi (x) \, \varphi (2y)$ and $\varphi (2x) \, \varphi (y)$,
and $\RR^{(3b)}$ only contributions of the types $\varphi (x)$ and $\varphi (y)$.
There are two corresponding parts of $\bb^{(3)}$, that is $\bb^{(3a)}$ and $\bb^{(3b)}$,
which satisfy $\nab^2 \bb^{(3a)} = - 5 k_{\rm H}^2 \bb^{(3a)}$
and $\nab^2 \bb^{(3b)} = -  k_{\rm H}^2 \bb^{(3b)}$
and equations of type of \eq{a11}.
The structure of $\bb^{(3a)}$ implies $\ol{\uu \x \bb^{(3b)}} = \bzo$
Only $\bb^{(3b)}$, for which \eq{a11} applies with $\hat{\bb}$ and $\hat{\RR}$
replaced by $\bb^{(3b)}$ and $\RR^{(3b)}$, respectively, and $k = 0$,
contributes to $\ol{\uu \x \bb^{(3b)}}$.

Detailed calculations along these lines deliver us
\EQA
\meanemf^{(4)}_x \!\!\! &=& \!\!\! \ol{u_y b^{(3)}_z} - \ol{u_z b^{(3)}_y}
    = u_0 \Rm^3 \big( a_{xx} \meanB_x + a_{xy} \meanB_y \big)
\nonumber\\
\meanemf^{(4)}_y \!\!\! &=& \!\!\! \ol{u_z b^{(3)}_x} - \ol{u_x b^{(3)}_z}
    = u_0 \Rm^3 \big( a_{yx} \meanB_x  + a_{yy} \meanB_x \big)
\label{b03}\\
\meanemf^{(4)}_z \!\!\! &=& \!\!\! \ol{u_x b^{(3)}_y} - \ol{u_y b^{(3)}_x} = 0 \, .
\nonumber
\ENA
with
\EQA
a_{xx} \!\!\! &=& \!\!\! - (\eta k_{\rm H}^2)^3 \big\{ \mbox{sc} [ \Gamma (\mbox{ss}, \mbox{ss}, \mbox{cc})
    + \Gamma (\mbox{cs}, \mbox{cs}, \mbox{sc}) ]
\nonumber\\
&& \qquad \qquad \quad + \mbox{cc} (\Gamma (\mbox{ss}, \mbox{ss}, \mbox{cc})
    + \Gamma (\mbox{cs}, \mbox{cs}, \mbox{cc}) ] \big\}
\nonumber\\
a_{xy} \!\!\! &=& \!\!\! - (\eta k_{\rm H}^2)^3 \big\{ \mbox{sc} [ \Gamma (\mbox{ss}, \mbox{cc}, \mbox{cs})
    - \Gamma (\mbox{cs}, \mbox{cc}, \mbox{ss}) ]
\nonumber\\
&& \qquad \qquad \quad - \mbox{cc} [\Gamma (\mbox{ss}, \mbox{sc}, \mbox{cs})
    - \Gamma (\mbox{cs}, \mbox{sc}, \mbox{ss}) ] \big\}
\nonumber\\
a_{yx} \!\!\! &=& \!\!\! + (\eta k_{\rm H}^2)^3 \big\{ \mbox{cs} [ \Gamma (\mbox{sc}, \mbox{ss}, \mbox{cc})
    - \Gamma (\mbox{cc}, \mbox{ss}, \mbox{sc}) ]
\label{b05}\\
&& \qquad \qquad \quad - \mbox{cc} [\Gamma (\mbox{sc}, \mbox{cs}, \mbox{cc})
    - \Gamma (\mbox{cc}, \mbox{cs}, \mbox{sc}) ] \big\}
\nonumber\\
a_{yy} \!\!\! &=& \!\!\! - (\eta k_{\rm H}^2)^3 \big\{ \mbox{cs} [ \Gamma (\mbox{sc}, \mbox{sc}, \mbox{cs})
    + \Gamma (\mbox{cc}, \mbox{cc}, \mbox{cs}) ]
\nonumber\\
&& \qquad \qquad \quad + \mbox{ss} [\Gamma (\mbox{sc}, \mbox{sc}, \mbox{ss})
    + \Gamma (\mbox{cc}, \mbox{cc}, \mbox{ss}) ] \big\}
\nonumber
\ENA
where
\EQA
\Gamma (f, g, h) &=&
\nonumber\\
&& \!\!\!\!\!\!\!\!\!\!\!\!\!\!\!\!\!\!\!\!\!\!\!\!\!\!\!\!\!\!\!\!\!\!\!\!\!
\int_0^\infty \!\!\! \int_0^\infty \!\!\! \int_0^\infty
    f (t -t') g (t - t' - t'') h (t - t' - t'' - t''')
\label{b07}\\
&&  \exp [- \eta k_{\rm H}^2 (t' + 2 t'' + t''')]
    \, \dd t' \, \dd t'' \dd t''' \, .
\nonumber
\ENA
The combinations of trigonometric functions in \eq{b05} can be expressed by the $\mbox{CC}$, $\mbox{CS}$,
$\mbox{SC}$ and $\mbox{SS}$ defined in \eq{eq37} and \eq{eq67}.
In this way we arrive at the results \eq{eq63} and \eq{eq65}.

\end{document}